\documentclass[superscriptaddress,groupedaddress,nofootnoteinbib,12pt]{article}
\usepackage{color}
\usepackage{graphicx}
\usepackage{dcolumn}
\usepackage{bm}
\usepackage{amssymb}
\usepackage{amsmath}
\usepackage{sectsty}
\usepackage{colortbl}
\usepackage{latexsym}
\usepackage{float}
\usepackage{ifthen}
\usepackage{caption,subfig}
\usepackage{enumerate}
\usepackage{url}
\usepackage{caption,subfig}	
\usepackage{jcappub}

\newcommand{\eff}{{\rm eff}}

\def\be{\begin{equation}}
\def\ee{\end{equation}}
\def\ba{\begin{eqnarray}}
\def\ea{\end{eqnarray}}
\def\beq{\begin{eqnarray}}
\def\eeq{\end{eqnarray}}

\def\mpl{M_{\rm Pl}}

\def\K{{\cal K}}

\def\L*{{\cal L}_*}
\def\L{\mathcal{L}}
\def\({\left(}
\def\){\right)}

\def\<{\langle}
\def\>{\rangle}

\def\cs2{c_{s}^{2}}

\def\be{\begin{equation}}
\def\ee{\end{equation}}
\def\ba{\begin{eqnarray}}
\def\ea{\end{eqnarray}}
\def\beq{\begin{eqnarray}}
\def\eeq{\end{eqnarray}}

\def\mpl{M_{\rm Pl}}

\def\K{{\cal K}}

\def\L*{{\cal L}_*}
\def\L{\mathcal{L}}
\def\({\left(}
\def\){\right)}

\def\<{\langle}
\def\>{\rangle}

 \def\be   {\begin{equation}}   \def\ee   {\end{equation}}

 \def\ba  {\begin{eqnarray}}   \def\ea  {\end{eqnarray}}

\renewcommand{\thesubsection}{\arabic{section}.\arabic{subsection}}

\begin{document}

\title{Cosmology in massive gravity with effective composite metric}

\author{Lavinia Heisenberg$^{a}$, Alexandre Refregier$^{b}$}
\affiliation{$^{a}$Institute for Theoretical Studies, ETH Zurich\\
Clausiusstrasse 47, 8092 Zurich, Switzerland}
\affiliation{$^{b}$Institute for Astronomy, Department of Physics, ETH Zurich,\\
Wolfgang-Pauli-Strasse 27, 8093, Zurich, Switzerland}

\emailAdd{lavinia.heisenberg@eth-its.ethz.ch}
\emailAdd{alexandre.refregier@phys.ethz.ch}

\abstract{This paper is dedicated to scrutinizing the cosmology in massive gravity. A matter field of the dark sector is coupled to an effective composite metric while a standard matter field couples to the dynamical metric in the usual way. For this purpose, we study the dynamical system of cosmological solutions by using phase analysis, which provides an overview of the class of cosmological solutions in this setup. This also permits us to study the critical points of the cosmological equations together with their stability. We show the presence of stable attractor de Sitter critical points relevant to the late-time cosmic acceleration. Furthermore, we study the tensor, vector and scalar perturbations in the presence of standard matter fields and obtain the conditions for the absence of ghost and gradient instabilities.
Hence, massive gravity in the presence of the effective composite metric can accommodate interesting dark energy phenomenology, that
can be observationally distinguished from the standard model according to the expansion history and cosmic growth.}

\maketitle

\section{Introduction}
The Standard Model of particle physics unifies the electro-weak and strong interactions and consists of elementary and composite particles described by the framework of quantum field theory. Particles correspond to the excited states of the underlying physical field. The naturally occurring ones are massive and massless spin-$0$, $1$, $1/2$ and $2$ particles. 
The construction of a mass term for a spin-$0$ particle is trivial since it does not alter the number of propagating physical degrees of freedom due to the absence of gauge symmetry to be broken. So far the only observed fundamental spin-$0$ particle in nature is the Higgs boson. Since spin-$0$ particles preserve the isotropy of the universe, they also could be a natural candidate for dark energy. One can allow for non-trivial self-interactions of the spin-$0$ particle, which on the other hand will have important consequences in multifaceted applications \cite{Horndeski:1974wa,Nicolis:2008in,Deffayet:2009wt,Deffayet:2009mn,DeFelice:2010nf,deRham:2011by,Burrage:2011cr,Heisenberg:2014kea}. Nature comprises also abelian and non-abelian spin-$1$ fields. Since these fields contain a gauge symmetry, breaking it alters the number of propagating degrees of freedom. Its dynamics can be captured by the Proca action, but there could also be non-trivial self-interactions \cite{Horndeski:1976gi,EPU10,PhysRevD.80.023004,BeltranJimenez:2013fca,Jimenez:2013qsa,Jimenez:2014rna,Heisenberg:2014rta,Tasinato:2014eka,Allys:2015sht,DeFelice:2016cri,Jimenez:2016isa,Chagoya:2016aar}.

The existence of a spin-$2$ particle in nature has not been directly observed yet. There are indirect observations through gravitational interactions that indicates the carrier of gravitational interactions to be a spin-$2$ particle. The recent observation of gravitational waves has imported more evidence for this to be the case \cite{Abbott:2016blz}. An important fundamental question is whether this spin-$2$ particle has exactly zero mass 
or eventually a small but non-zero mass. The recent observation of gravitational waves does not put tight constraints on the mass of the graviton, at least not as tight as the already existing ones. The construction of the unique mass term at the linear level was done already in the 1940's by Fierz and Pauli without introducing an additional ghost degree of freedom at the classical level \cite{Fierz:1939ix}. In contrast to the standard Proca field, taking the zero mass limit gives a discrete difference between the massless and massive theory \cite{vanDam:1970vg,Zakharov:1970cc}. This vDVZ discontinuity manifests itself in the graviton exchange amplitudes and the predictions of General Relativity can not be recovered in the limit of massless gravitons. The discontinuity can be better understood by writing the interactions in terms of the St\"uckelberg fields and restoring the diffeomorphism invariance. By doing that, one immediately observes that the scalar part of the massive graviton couples to the trace of the energy momentum tensor and causes the discontinuity. Interestingly, the discontinuity is absent on Anti-de Sitter backgrounds \cite{Kogan:2000uy, Porrati:2000cp}, even though this property might be only true at the tree level \cite{Dilkes:2001av}. Very soon, it was realized by Vainshtein, that this discontinuity is just an artifact of the linear approximation and the non-linear interactions become appreciable on small scales \cite{Vainshtein:1972sx}. Thus, in order to recover the predictions of General Relativity, the non-linear completion of the theory was necessary. However, these non-linear extensions usually reintroduce the Boulware-Deser ghost instability \cite{BoulwareD}. Nevertheless, it was possible to circumvent this seemingly no-go result and constitute a unique ghost-free non-linear theory of massive gravity \cite{deRham:2010ik,deRham:2010kj,Hassan:2011vm,Hassan:2011hr}. \\

One important and interesting consequence of a non-zero mass of the graviton is its effect on cosmological scales. Its phenomenological application is diverse and rich. Even if its standard formulation faces a theoretically well expected no-go result for flat FLRW solutions in the case of flat fiducial metric \cite{PhysRevD.84.124046}, its extensions beyond the standard formulation might yield interesting phenomenology. Starting with self-accelerating open FLRW solution, one soon realized that this excites non-linear ghost instability \cite{PhysRevLett.109.171101, DeFelice:2013awa}. This negative result has motivated the study of more general fiducial metrics, which unfortunately either suffered from Higuchi type instabilities for de Sitter reference metric
\cite{Fasiello:2012rw,Langlois:2012hk} or from the absence of acceleration for anti-de Sitter reference metric \cite{MartinMoruno:2013gq}. It seems that the destiny of the self-accelerating branch solutions does not depend much on the choice of the reference metric and is doomed to have the same instability as the open solutions \cite{Gumrukcuoglu:2011zh,Gumrukcuoglu:2012wt}.
One can of course consider further extensions of the theory by either adding new degrees of freedom \cite{Huang:2012pe,PhysRevD.87.064037,DeFelice:2013tsa} or breaking some of the underlying symmetries \cite{deRham:2014gla,DeFelice:2015yha}.  \\

In the context of the quantum stability of the theory \cite{deRham:2012ew,deRham:2013qqa}, a new and unexplored branch of research along the line of consistent matter couplings has been proposed in \cite{deRham:2014naa}. At the classical level the potential interactions were tuned in a specific way to guarantee the absence of the Boulware-Deser ghost. The requirement to maintain this property also at the quantum level severely restricts the allowed matter coupling. The difficulties encountered in the cosmological application mentioned above arise only if one restricts the matter fields to couple minimally to one of the metrics. One gains a richer phenomenology if one gives up this restriction. In fact, non-minimal matter couplings through
 a very specific effective composite metric built out of the dynamical and fiducial metric offer promising cosmological solutions, even though they usually reintroduce the Boulware-Deser ghost\cite{deRham:2014naa, deRham:2014fha}. However, in the metric language there is a unique effective metric that keeps the theory ghost-free up to the strong coupling scale \cite{Huang:2015yga,Heisenberg:2015iqa}. This allows us to consider these couplings as a consistent effective field theory with the cut-off equal to or lower than the mass of the ghost. In the unconstrained Vielbein formulation of the theory, one can construct other types of effective vielbeins that are free of the Boulware-Deser ghost up to the strong coupling scale \cite{Melville:2015dba}.
Even if one can construct these effective field theories, it would be more desirable to preserve the ghost freedom fully non-linearly. There was the hope that the unconstrained vielbein formulation might accomplish this requirement \cite{Hinterbichler:2015yaa}, which also was soon disproven \cite{deRham:2015cha}. However, one can maintain the ghost freedom fully non-linearly in the partially constrained Vielbein formulation at the price of losing local Lorentz symmetry \cite{DeFelice:2015yha}. Further phenomenological consequences of matter couplings in massive (bi-)gravity were studied in \cite{Khosravi:2011zi, Akrami:2012vf,Akrami:2013ffa,Tamanini:2013xia,Akrami:2014lja,Yamashita:2014fga,Noller:2014sta,Schmidt-May:2014xla,Enander:2014xga,Solomon:2014iwa,Soloviev:2014eea,Heisenberg:2014rka,Heisenberg:2015wja,Blanchet:2015sra,Blanchet:2015bia,Bernard:2015gwa,Mukohyama:2014rca,Lagos:2015sya,Matas:2015qxa,DeFelice:2015yha,Gao:2016vtt,DeFelice:2016tiu}. \\

In this work, we will follow the preliminary works on the cosmological application of massive gravity in the presence of matter fields that couple to the effective metric \cite{deRham:2014naa,Solomon:2014iwa,Gumrukcuoglu:2014xba}. As mentioned before, if one of the matter fields couple to the effective metric, then the no-go result for the existence of exact FLRW solutions in massive gravity can be avoided \cite{deRham:2014naa}, which can yield promising cosmological solutions without resorting to any new additional degrees of freedom besides the ones of massive gravity. As was argued in \cite{deRham:2014naa}, these doubly coupled matter fields are expected to be part of a dark sector and should not be considered as fields that drive the cosmic expansion of the Universe. A first look to the Friedmann equation makes this statement quite transparent. The matter fields living on the effective metric enter the Friedmann equation in a peculiar way and hence contribute in a non-standard way to the evolution of the Universe. Following this, we shall assume that all the standard matter fields like dust and radiation still couple only to the dynamical $g$ metric. We can describe the standard matter fields that only couple to the dynamical metric as standard perfect fluids. However, for the doubly coupled matter fields coming from the dark sector, special attention is needed and one can not apply the same standard perfect fluid approach. For that reason we shall keep a field description of the matter fields following \cite{deRham:2014naa}. For the purpose of obtaining a general overview of the class of cosmological solutions that is expected to be encountered in our setup, we shall perform a dynamical system analysis. Not only can we obtain all of the existing critical points of the cosmological equations in this way, but also their stability. 

We will first review massive gravity in the presence of doubly coupled matter fields in Section \ref{sec:dRGT}. After working out the underlying background equations of our setup in Section \ref{sec:background_evolution}, we will first study the system at late times when the standard matter fields are subdominant in Section \ref{sec:late_time} and move on to the full dynamical system analysis of cosmological solutions in Section \ref{sec:general}. We will show the presence of stable de Sitter critical points, which yields acceleration at late times and can play the role of dark energy. The complete set of equations of the autonomous system is illustrated for a special subclass of the model parameters in Section \ref{sec:alpha2}. Finally, we will study the stability of the tensor, vector and scalar perturbations on top of the background in Section \ref{sec:perturbations} and conclude in Section \ref{sec:conclusion}.


\section{Massive gravity with doubly coupled matter}
\label{sec:dRGT}
We first review the allowed interactions in the theory of massive gravity and setup our framework. We consider the action for massive gravity and an additional matter field $\chi$ that couples to the composite effective metric proposed in \cite{deRham:2014naa}. Furthermore, we shall assume that the ordinary matter fields still couple minimally to the physical metric $g$. This coupling will be represented by an additional field $\phi$. Thus, our action reads

\begin{equation}\label{action_MG_effcoupl}
\mathcal{S} = \int \mathrm{d}^4x \big[ \frac{\mpl^2}{2} \sqrt{-g}\left(R[g]-\frac{m^2}{2}\sum_n \alpha_n{\cal U}_n[\cal K]  \right) +\mathcal{L}_\chi(g_{\rm eff},\chi)+\mathcal{L}_{\rm matter}(g,\phi)\big]\,,
\end{equation}
with the precise allowed potential interactions for the massive graviton given by \cite{deRham:2010ik,deRham:2010kj}
\begin{eqnarray}
\mathcal{U}_0[\mathcal{K}] &=& \mathcal{E}^{\mu\nu\rho\sigma}  \mathcal{E}_{\mu\nu \rho\sigma} = 24, \nonumber\\
\mathcal{U}_1[\mathcal{K}] &=& \mathcal{E}^{\mu\nu\rho\sigma}  \mathcal{E}^{\alpha}_{\;\;\;\nu \rho\sigma} \K_{\mu\alpha} = 6 [\K], \nonumber\\
\mathcal{U}_2[\mathcal{K}] &=& \mathcal{E}^{\mu\nu\rho\sigma}  \mathcal{E}^{\alpha\beta}_{\;\;\;\;\; \rho\sigma} \K_{\mu\alpha} \K_{\nu\beta} = 2\left( [\K]^2-[\K^2]\right), \nonumber\\
\mathcal{U}_3[\mathcal{K}] &=& \mathcal{E}^{\mu\nu\rho\sigma}  \mathcal{E}^{\alpha\beta\kappa}_{\;\;\;\;\;\;\; \sigma} \K_{\mu\alpha} \K_{\nu\beta}  \K_{\rho\kappa}=[\K]^3-3[\K][\K^2]+2[\K^3],  \nonumber\\
\mathcal{U}_4[\mathcal{K}] &=& \mathcal{E}^{\mu\nu\rho\sigma}  \mathcal{E}^{\alpha\beta\kappa\gamma} \K_{\mu\alpha} \K_{\nu\beta}  \K_{\rho\kappa}  \K_{\sigma\gamma} =[\K]^4-6[\K]^2[\K^2]+3[\K^2]^2+8[\K][\K^3]-6[\K^4]\,.
\end{eqnarray}
The presence of the Levi-Cevita tensor $\mathcal{E}$ is crucial for the quantum stability of the interactions \cite{deRham:2012ew, deRham:2013qqa}. The fundamental tensor of the theory $\K$ consists of couplings between the dynamical metric $g$ and the reference metric $f$ in form of a square root
\begin{equation}
\K^\mu _\nu[g,f] =\delta^\mu_\nu - \left(\sqrt{g^{-1}f}\right)^\mu_\nu \,.
\end{equation}
The pure Einstein-Hilbert kinetic term is diffeomorphism invariant. The inclusion of the potentials in \ref{action_MG_effcoupl} breaks this invariance. However, gauge symmetries are redundancies of description and can be easily restored by introducing redundant variables. In order to restore back the gauge symmetry in our action \ref{action_MG_effcoupl} we can use the St\"uckelberg trick and promote the Minkowski reference metric to the space-time tensor by including the four St\"uckelberg fields $S^a$
\begin{equation}\label{Stueckelbergfields}
f_{\mu\nu} = \eta_{ab}\partial_\mu S^a \partial_\nu S^b\,.
\end{equation}
The unitary gauge corresponds to $S^a=x^a$. Based on studies of quantum stability of the theory, a new effective composite coupling built out of the two metrics in a very specific way was proposed in \cite{deRham:2014naa}. This coupling to both metrics ensures that one-loop corrections from virtual matter fields do not destroy the special structure of the potentials. The price to pay is the loss of the naturalness argument.
In our action, it is the matter Lagrangian $\mathcal{L}_\chi$ that lives on the effective metric 
\begin{equation}
g^\eff_{\mu\nu} \equiv \alpha^2 g_{\mu\nu}+2\,\alpha\,\beta\, g_{\alpha\mu} \left(\sqrt{g^{-1}f}\right)^\alpha_{\nu} + \beta^2 f_{\mu\nu}\,.
\label{eq:geff}
\end{equation}
This effective metric is unique in the sense that the Boulware-Deser degree of freedom is not generated in the decoupling limit of the theory and its volume element  
\begin{equation}
\label{eq:detgeff}
\sqrt{-g_{\eff}}= \sqrt{-g}\  \det\(\alpha +\beta\left(\sqrt{g^{-1}f}\right)^\mu_\nu\)\,,
\end{equation}
corresponds exactly to a particular choice of the allowed potential interactions that can be compactly written as the expansion of a deformed determinant \cite{Hassan:2011vm}
\begin{equation}
\label{eq:detgeff2}
\sqrt{-g_{\eff}}= \sqrt{-g}\  \sum_{n=0}^4 \frac{(-\beta)^n}{n!}(\alpha+\beta)^{4-n} \mathcal{U}_n[K]\,.
\end{equation}
Without loss of generality, we will consider a generic scalar field $\chi$ that minimally couples to the effective metric for simplicity
\begin{equation}
\mathcal{L}_{\chi} =\sqrt{-g_\text{eff}}\,P(\chi, X_\chi)\,,
\label{eq:chiaction}
\end{equation}
with $X_\chi$ standing for the standard kinetic term of the $\chi$ field
\begin{eqnarray}
  X_\chi \equiv -g_\text{eff}^{\mu\nu}\partial_{\mu}\chi\partial_{\nu}\chi
\,.
\label{matter}
\end{eqnarray}
The corresponding energy density, pressure and sound speed of the $\chi$ field can be expressed as
\begin{equation}\label{densities_pressure_cs}
 \rho_\chi \equiv 2\partial_{X_\chi}P(\chi, X_\chi)X_\chi - P(\chi, X_\chi), \;\;\;  P_\chi \equiv P(\chi, X_\chi), \;\;\;
 c_\chi^2 \equiv \frac{\partial_{X_\chi}P(\chi, X_\chi)}{2\partial^2_{X_\chi}P(\chi, X_\chi)X_\chi+\partial_{X_\chi}P(\chi, X_\chi)}\,.
\end{equation}  
Note, that the $\chi$ field is not a standard matter field and presumably belongs to a dark sector. Its presence ensures the existence of flat FLRW solutions \cite{deRham:2014naa}. We will also include standard matter fields that live in the standard metric for phenomenological viability
\begin{equation}
\mathcal{L}_{\rm matter} =\sqrt{-g}\,\tilde{P}(\phi, X_\phi)\,,
\label{eq:chiaction}
\end{equation}
with this time $ X_\phi \equiv -g^{\mu\nu}\partial_{\mu}\phi\partial_{\nu}\phi$. The corresponding energy density, pressure and sound speed are defined accordingly as in equation \ref{densities_pressure_cs} but for $\tilde{P}(\phi, X_\phi)$.
We will also set the tadpole $\mathcal{U}_1$ and cosmological constant $\mathcal{U}_0$ contributions in \ref{action_MG_effcoupl} to zero, since we are interested in the self-accelerating solutions. This constitutes our model that we will study in detail in this work. First we shall work out the equations that dictate the background dynamics in next section.


\section{Background equations}\label{sec:background_evolution}
Very early on it was showed that massive gravity in its original formulation is subject to a no-go theorem for flat FLRW solutions \cite{PhysRevD.84.124046}. The equation of motion for the St\"uckelberg field gives the constraint that the scale factor can not evolve in time. The matter coupling through the effective metric modifies the St\"uckelberg field equation of motion and one can construct exact FLRW solutions with flat reference metric \cite{deRham:2014naa}. In this work, we follow \cite{deRham:2014naa,Solomon:2014iwa,Gumrukcuoglu:2014xba} and investigate in more detail the cosmological background evolution in the presence of the standard matter fields. As usual, we shall assume an homogeneous and isotropic flat FLRW ansatz for the dynamical metric
\begin{equation}
ds_g^2=-N^2 dt^2 +a^2 \delta_{ij} dx^idx^j\,,
\end{equation}
while, for the fiducial metric, we consider a pull-back of the Minkowski metric in the St\"uckelberg field space to the physical space-time
\begin{equation}
ds_f^2= f_{\mu\nu}dx^\mu dx^\nu = -\dot{f}^2dt^2 + a_0^2 \delta_{ij} dx^idx^j\,.
\label{eq:reference}
\end{equation}
This means that we have chosen $S^0=f(t)$, $S^i = a_0 x^i$ for the St\"uckelberg fields. The unitary gauge $S^a=x^a$ corresponds to $f(t)=t$ and $a_0=1$. For an homogeneous and isotropic background the effective metric defined in equation (\ref{eq:geff}) takes the form
\begin{equation}
ds^2_\eff = -N^2_\eff dt^2+a_\eff^2 \delta_{ij}dx^idx^j\,,
\end{equation}
with $N_{\eff}$ and $a_{\eff}$ being the effective lapse and scale factor respectively
\begin{equation}
N_\eff \equiv \alpha\,N+\beta\,\dot{f}\,,\qquad
a_\eff \equiv \alpha\,a+\beta\,a_0\,.
\end{equation}
Furthermore, we also assume that the background matter fields also depend only on time $\chi=\chi(t)$ and $\phi=\phi(t)$. Our action (\ref{action_MG_effcoupl}) in the mini-superspace becomes (up to total derivatives):
\begin{eqnarray}
\frac{S}{V} &=& \mpl^2\int dt \,a^3 N\,\Bigg\{-3H^2-m^2\left[\rho_m+rAQ\right]\Bigg\} \nonumber\\
&&+\int dt \,a_\eff^3 N_\eff P(\chi, X_\chi)+\int dt \,a^3 N \tilde{P}(\phi, X_\phi) \,,
\label{eq:minisuperspace}
\end{eqnarray}
where we have defined the following quantities for our convenience
\begin{eqnarray}\label{shortcuts}
A\equiv a_0/a\,,\qquad H\equiv \frac{\dot{a}}{a}\,,\qquad
r\equiv \frac{\dot{f}/a_0}{N/a}\,, \\
\rho_m (A)\equiv U(A)-\frac{A}{4}\, \partial_AU \,,\qquad
Q(A) \equiv \frac{1}{4}\partial_AU\,,
\end{eqnarray}
with $U(A)  \equiv 6\,\sum_{n=2}^4\,\alpha_n (1-A)^n$ and $A$ denoting the ratio of the scale factors, $H$ the expansion rate of the physical $g$ metric, $r$ the speed of light
propagating in the $f$ metric in the units of the one propagating in the $g$ metric and $\rho_m$ the dimensionless effective energy density from the mass term. 
Next we compute the background equations of motion by varying the action (\ref{eq:minisuperspace}) with respect to $N$, $a$, $\chi$, $\phi$ and $f$. First, we vary the action (\ref{eq:minisuperspace}) with respect to the lapse $N$ to obtain the Friedmann equation 
\begin{equation}
3\,\frac{H^2}{N^2} =  m^2 \rho_m +\frac{\rho_\phi}{\mpl^2}+\frac{\alpha\,a_\eff^3}{\mpl^2\,a^3}\rho_\chi\,.
\label{eq:eqN}
\end{equation}
Then, we vary the mini-superspace action (\ref{eq:minisuperspace}) with respect to the scale factor $a$ and combine it with the Friedmann equation to obtain the acceleration equation 
\begin{equation}
\frac{2\,\dot{H}}{N^2}=\frac{2H\dot{N}}{N^3}+ m^2\,J\,A\,(r-1)-\frac{\rho_\phi+P_\phi}{\mpl^2} - \frac{\alpha\,a_\eff^3}{\mpl^2a^3}\left[
\rho_\chi + \frac{N_\eff/a_\eff}{N/a} P_\chi \right]\,,
\label{eq:eqa}
\end{equation}
with $J=\frac13\partial_A\rho_m(A)$. The matter fields equations of motion are just the standard conservation equation,
\begin{eqnarray}
\frac{1}{N_\eff}\,\dot{\rho}_\chi+3\,\frac{H_\eff}{N_\eff}\,(\rho_\chi+P_\chi)&=&0\,,  \nonumber\\
\frac{1}{N}\,\dot{\rho}_\phi+3\,\frac{H}{N}\,(\rho_\phi+P_\phi)&=&0\,.
\label{eq:eqchi}
\end{eqnarray}
Finally, the equation of motion for the St\"uckelberg field yields
\begin{equation}
m^2\,\mpl^2J=\frac{\alpha\beta\,a_\eff^2}{a^2} P_\chi\,.
\label{eq:eqf}
\end{equation}
The system of equations of motion are related by the contracted Bianchi identity,
\begin{equation}
\frac{\partial}{\partial t} \frac{\delta S}{\delta N} - \frac{\dot{a}}{N}\frac{\delta S}{\delta a}- \frac{\dot{\chi}}{N}\frac{\delta S}{\delta \chi}- \frac{\dot{\phi}}{N}\frac{\delta S}{\delta \phi}- \frac{\dot{f}}{N}\frac{\delta S}{\delta f} =0\,.
\label{eq:bianchi}
\end{equation}
Our goal is to study these background equations in detail and investigate whether the system admits interesting de Sitter critical points despite the presence of the matter fields. The presence of de Sitter critical points will be important for the desired dark energy phenomenology.


\section{Late time behaviour}\label{sec:late_time}
In this section we first investigate the solutions with $\dot{H}=0$, or equivalently $\ddot{A}=0$ at late times. We assume that the matter field that couples only to the physical metric $g$ is subdominant at very late times when dark energy dominates, so we impose $P_\phi=0$ and $\rho_\phi=0$. We first use the constraint equation \ref{eq:eqf} to solve for $P_\chi$
\begin{equation}
P_\chi= \frac{m^2\mpl^2 J}{\alpha\beta(\alpha+\beta A)^2} \,.
\end{equation}
From the acceleration equation \ref{eq:eqa} after replacing $P_\chi$ by the above expression we obtain that 
\begin{equation}
\dot{A}=\pm \frac{A}{\sqrt{2}}\left(-\frac{m^2(\alpha+\beta A)J}{\beta}-\frac{\alpha(\alpha+\beta A)^3\rho_\chi}{\mpl^2}\right)^{\frac12}  \,,
\end{equation}
and the equation of the motion for the matter field that lives on the $g_{\eff}$ metric yields
\begin{equation}
 \dot{\rho}_\chi= \frac{\pm3\mpl^2}{\sqrt{2}(\alpha+\beta A)^4}\left(-\frac{m^2(\alpha+\beta A)J}{\beta}-\frac{\alpha(\alpha+\beta A)^3\rho_\chi}{\mpl^2}\right)^{3/2} \,.
\end{equation}
We can further substitute the expressions for $\dot{A}$ and $P_\chi$ into the Friedmann equation to obtain the expression for $\rho_\chi$
\begin{equation}
\rho_\chi= \frac{-m^2\mpl^2(3(\alpha+\beta A) J+2\beta \rho_m}{5\alpha\beta(\alpha+\beta A)^3} \,.
\end{equation}
Finally, we use the expression for $\rho_\chi$ in order to express the equation for $\dot{A}$ as 
\begin{eqnarray}\label{auto_dS}
\dot{A}&=& \pm \frac{A}{\sqrt{5}} \left(\frac{m^2}{\beta}(-(\alpha+\beta A) J+\beta \rho_m)\right)^{\frac12}
\end{eqnarray}
We can solve the equation for $\dot{A}=-a_0 \dot{a}/a^2$ to find the evolution for the scale factor (for simplicity we put $a_0=1$) 
\begin{equation}
H=\frac{\dot{a}}{a}=\frac{\pm1}{\sqrt{10\beta a^2}}\left(m^2(\beta\kappa_2-2\alpha\kappa_3-a(-4\beta\kappa_1+2\alpha\kappa_2+(2\alpha\kappa_1+\beta(6\kappa_1+3\kappa_2+2\kappa3))a))\right)^{\frac12} \,,
\end{equation}
where we introduced new combinations of the parameters
\begin{equation}
\kappa_1=3(\alpha_2+\alpha_3)+\alpha_4\,,\quad 
\kappa_2=-2 (\alpha_2+2\,\alpha_3+\alpha_4)\,,\quad 
\kappa_3 =\alpha_3+\alpha_4\,.
\end{equation}
The solution to this equation is simply given by
\begin{eqnarray}
a&=&\left( e^{\frac{\pm m\sqrt{-\tilde{\kappa}_1}(\sqrt{10}t+10c)}{10\sqrt{\beta}}} +4\beta\kappa_1-2\alpha\kappa_2\right)(2\tilde{\kappa}_1)^{-1}\nonumber\\
&+& \left(e^{-\frac{\pm m\sqrt{-\tilde{\kappa}_1}(\sqrt{10}t+10c)}{10\sqrt{\beta}}} (\beta^2\tilde{\kappa}_2-2\alpha\beta\tilde{\kappa}_2+\alpha^2(\kappa_2^2-4\kappa_1\kappa_3) )\right)(2\tilde{\kappa}_1)^{-1} \,,
\end{eqnarray}
where $c$ is the integration constant and we further introduced the shortcut notations for convenience 
\begin{equation}
\tilde{\kappa}_1=2\alpha\kappa_1+\beta(6\kappa_1+3\kappa_2+2\kappa_3)\,,\quad 
\tilde{\kappa}_2= 4\kappa_1^2+6\kappa_1\kappa_2+3\kappa_2^2+2\kappa_2\kappa_3 \,,\quad 
\tilde{\kappa}_3 = \kappa_1(\kappa_2+6\kappa_3)+3\kappa_2\kappa_3+2\kappa_3^2 \,.
\end{equation}
As next we can compute the critical points of our autonomous system at late times. From the vanishing of $\dot{A}$ in \ref{auto_dS}, we obtain
\begin{equation}
m^2(\beta\kappa_2-2\alpha\kappa_3-a(-4\beta\kappa_1+2\alpha\kappa_2+\tilde{\kappa}_1a))=0
\end{equation}
which has the two solutions as the critical points
\begin{equation}
a_{c}=\frac{1}{\tilde{\kappa}_1}\left( 2\beta\kappa_1-\alpha\kappa_2\pm\frac12\left((4\beta\kappa_1-2\alpha\kappa_2)^2+4\tilde{\kappa}_1(\beta\kappa_2-2\alpha\kappa_3)\right)^{\frac12} \right)
\end{equation}
These solutions satisfy automatically the vanishing of $ \dot{\rho}_\chi$ in equation \ref{auto_dS}. At the critical points we have further that the energy density and the
pressure of the matter field are given by
\begin{eqnarray}
\rho_\chi&=&-\frac{m^2\mpl^2(3a_c^2(3\beta\kappa_1+\alpha\kappa_2)+5\beta\kappa_3+3a_c(2\beta\kappa_2+\alpha\kappa_3)}{5\alpha\beta(a_c\alpha+\beta)^3}  \nonumber \\
&-&\frac{a_c^3(3\alpha\kappa_1-\beta(6\kappa_1+3\kappa_2+2\kappa_3)))}{5\alpha\beta(a_c\alpha+\beta)^3} \,, \nonumber \\
P_\chi&=&\frac{m^2\mpl^2(a_c(a_c\kappa_1+\kappa_2)+\kappa_3)}{\alpha\beta(a_c\alpha+\beta)^2} \,.
\end{eqnarray}
For the stability around the critical points the linearized system evaluated at the critical points
\begin{equation}
\frac{d}{dt} (\delta A)\big|_{a_c} = \lambda \delta A)\big|_{a_c} \,,
\end{equation}
requires the following condition
\begin{equation}
\lambda=m^2\tilde{\kappa}_1\left((4\beta\kappa_1-2\alpha\kappa_2)^2+4\tilde{\kappa}_1(\beta\kappa_2-2\alpha\kappa_3)\right)^{\frac12} <0 \,,
\end{equation}
to be satisfied. Furthermore, we have to impose that $\tilde{\kappa}_1<0$ in order for the solutions to be real. In the next section, we investigate the system beyond
the approximative assumption of the late-time asymptotic form of the expansion history, since the late-time asymptotic solution does not correctly describe our current epoch, 
which is just in the transition between matter domination and accelerating expansion. We will abandon the restriction $\dot{H}=0$ and explore the presence of the matter field,
that lives in the standard space-time metric.


\section{The general case}\label{sec:general}
In this section, we study the dynamical system of cosmological solutions using phase analysis for the model including all the parameters and in the presence of the standard matter fields. 
The dynamical system analysis will allow us to obtain a general overview of the class of cosmological solutions that one can expect
to find in massive gravity with the effective coupling. This will not only give us the critical points of the cosmological equations but also their stability.  
For the purpose of the dynamical system analysis we transform the equations to be analyzed into an autonomous system. We first avoid
the direct dependence on the scale factor using the constraint equation that arose from the St\"uckelberg equation
\begin{equation}
m^2\mpl^2(\kappa_1+A(\kappa_2+\kappa_3A))=\alpha\beta(\alpha+\beta A)^2P_\chi \,,
\end{equation}
which gives
\begin{equation}
A=\frac{2\alpha^2\beta^2P_\chi +m\mpl(-m\mpl\kappa_2+ \bar{P}_\chi)}{2m^2\mpl^2\kappa_3-2\alpha\beta^3P_\chi} \,,
\end{equation}
where further non-trivial dependence on the parameters of the theory is enclosed in the introduced variable $\bar{P}_\chi$
\begin{equation}
\bar{P}_\chi=\pm\left(m^2\mpl^2(\kappa_2^2-4\kappa_1\kappa_3)+4\alpha\beta(\beta^2\kappa_1-\alpha\beta\kappa_2+\alpha^2\kappa_3) P_\chi\right)^{\frac12}\,.
\end{equation}
We have two branches of solutions for $A$ depending on the sign of $\bar{P}_\chi$. We substitute the solution for $A$ back into the Friedmann equation and solve for $\rho_\chi$. This depends in a non-trivial way on $P_\chi$, $H$ and $\rho_\phi$. Without loss of generality, we further assume that $P_\phi=0$ for simplicity. Using the accelerating equation \ref{eq:eqa} we write $\dot{H}$ in terms of $H$ and $P_\chi$. After rewriting the energy density of the standard matter field in terms of $\Omega_\phi=\rho_\phi/(6\mpl^2H^2)$, we solve the corresponding conservation equation for $\dot{\Omega}_\phi$ in terms of $H$, $P_\chi$ and $\Omega_\phi$. Finally, we use the conservation equation of the matter field living on the effective metric to solve for $\dot{P}_\chi$ in terms of $H$, $P_\chi$ and $\Omega_\phi$. We can then reduce the equations to the following autonomous system
\begin{eqnarray} \label{auto_sys}
\frac{dH}{dN}&=&F_1(H,P_\chi) \nonumber\\
\frac{d\Omega_\phi}{dN}&=&F_2(H,P_\chi,\Omega_\phi)  \nonumber\\
\frac{dP_\chi}{dN}&=&F_3(H,P_\chi,\Omega_\phi)
\end{eqnarray}
where we used $dN=Hdt$ and the functions $F_i$ depend in a non-trivial way on the variables in the brackets (their exact dependence is given in the Appendix \ref{appendix}). In order to obtain the critical points, we have to solve the vanishing of the right hand side of the autonomous system \ref{auto_sys}. First of all, from $\frac{dH}{dN}=0$, we obtain the value for $H$ in terms of $P_\chi$. Substituting this expression in the vanishing of $\frac{d\Omega_\phi}{dN}$ immediately shows that all the critical points satisfy the vanishing of the standard matter field
\begin{equation}
\Omega_\phi=0 \qquad \;\; \;\; \text{at the critical points}\,.
\end{equation}
Finally, the vanishing of the third equation of the autonomous system $dP_\chi/dN=0$ yields
\begin{equation}\label{auto_eq3}
(P_\chi\alpha\beta^3-m^2\mpl^2\kappa_3)\bar{P}_\chi \frac{f_1(P_\chi)}{f_2(P_\chi)}=0
\end{equation}
where the ratio of the two functions $f_1$ and $f_2$ simplifies to give 
\begin{equation}
\frac{f_1(P_\chi)}{f_2(P_\chi)}= \frac{1}{m\mpl\alpha\beta(-\beta\kappa_2+2\alpha\kappa_3)} \,.
\end{equation}
From the vanishing of equation \ref{auto_eq3} we obtain the following values for the pressure 
\begin{eqnarray}
P^{(I)}_\chi=\frac{m^2\mpl^2\kappa_3}{\alpha\beta^3}\,, \qquad
P^{(II)}_\chi=-\frac{m^2\mpl^2(\kappa_2^2-4\kappa_1\kappa_3)}{4\alpha\beta(\beta^2\kappa_1-\alpha\beta\kappa_2+\alpha^2\kappa_3)}\,,\qquad
P^{(III)}_\chi=\frac{m^2\mpl^2\kappa_1}{\alpha^3\beta} \,.
\end{eqnarray}
We can substitute these values for the pressure obtained from vanishing $dP_\chi/dN=0$ into the expression for $H$ in order to find the critical points for $H$. The first value for the pressure $P^I_\chi$ requires special attention. Even though it makes $dP_\chi/dN$ vanish, it also makes $dH/dN$ diverge since the value of $H$ obtained from vanishing $dH/dN=0$ is inversely proportional to $(P_\chi\alpha\beta^3-m^2\mpl^2\kappa_3)$. The exact behaviour at this point should be investigated in more detail and in general will depend on the parameters of the theory. For this purpose, we consider a small deviation from $P^{(I)}_\chi$
\begin{equation}
P_\chi=P^{(I)}_\chi + \epsilon
\end{equation}
and consider the $\epsilon\to0$ limit. Around this point, the autonomous system scales as
\begin{eqnarray}
\frac{dH}{dN}&\approx&\frac{m^6\mpl^4(\beta\kappa_2-2\alpha\kappa_3)^3}{4\alpha^2\beta^9\epsilon^2}+\mathcal{O}(\epsilon^{-1}) \,, \nonumber\\
\frac{d\Omega_\phi}{dN}&\approx&-\frac{m^6\mpl^4(\beta\kappa_2-2\alpha\kappa_3)^3\Omega_\phi}{2\alpha^2\beta^9H\epsilon^2}+\mathcal{O}(\epsilon^{-1}) \,, \nonumber\\
\frac{dP_\chi}{dN}&\approx&H\epsilon+\mathcal{O}(\epsilon^{2}) \,.
\end{eqnarray}
Depending on the choice of the parameters, it can become a separatrix, but it does not represent a critical point. On the other hand, the second value $P^{(II)}_\chi$ plugged back into the expression for $H$ results in
\begin{equation}
H^{(II)}=\frac{\pm m}{\sqrt{6}} \left(-3\kappa_2-2\kappa_3+\frac{1}{\beta}\left(-2(\alpha+3\beta)\kappa_1+\frac{(-2\beta\kappa_1+\alpha \kappa_2)^2}{-\beta\kappa_2+2\alpha\kappa_3}\right)\right)^{\frac12} \,.
\end{equation}
This constitutes the first two non-trivial critical points of the system. They differ only by an overall sign in front of the square root. For the stability around these two critical points characterised by $(H^{(II)},P^{(II)}_\chi,\Omega_\phi=0)$ we shall investigate the linearized system
\begin{eqnarray}
\frac{d}{dt}\left(\begin{array}{c}
\delta H \\ \delta  \Omega \\ \delta P_\chi \end{array}\right)  &=& M^{(II)} \left(
\begin{array}{c}
\delta H \\\delta  \Omega \\ \delta P_\chi\end{array}\right) 
\end{eqnarray}
evaluated at the critical points with the corresponding matrix $M^{(II)}$ expressed as
\begin{eqnarray}
M^{(II)} =
\begin{pmatrix}
-3&0&m_{13} \\
0&-3&0 \\
0&0&m_{33} 
 \end{pmatrix} \,,
\end{eqnarray}
where the components $m_{13}$ and $m_{33}$ are inversely proportional to $\bar{P}_\chi$. The Eigenvectors of $M^{(II)}$ correspond to $(0,1,0)$, $(1,0,0)$ and $(\frac{m_{13}}{3+m_{33}},0,1)$. In order for the system to be stable, all the eigenvalues $\lambda_i$ have to be negative, since the system evolves as $e^{\lambda_i t}$ close to the critical points. The stability condition  actually corresponds to the convergence of nearby trajectories. In other words a stable critical point corresponds to an attractor. The eigenvalues of $M^{(II)}$ are simply given by 
\begin{equation}
\lambda_1=-3 \,, \qquad \lambda_2=-3 \,, \qquad \lambda_3=m_{33} \,.
\end{equation}
As one can see, for the stability condition, the sign of the component $m_{33}$ is important. We have to impose that it is negative in order to obtain an attractor de Sitter critical point. On closer inspection, one observes that
$m_{33}$ is of the form
\begin{equation}
m_{33}\sim \pm \frac{m^{10}\mpl^{10}(2\beta\kappa_1-\alpha \kappa_2)(\beta\kappa_2-2\alpha\kappa_3)^9}{128(\beta^2 \kappa_1 - \alpha \beta \kappa_2+\alpha^2\kappa_3)} \frac{1}{m\mpl \bar{P}_\chi (f_3(P_\chi))^2}
\end{equation}
with $f_3$ being a non-trivial function depending on $P_\chi$ and all the parameters of the theory. The specific form for $f_3$ is irrelevant for our analysis. The important observation to be made is that it comes squared and hence will not change the sign of the eigenvalue $\lambda_3$. In order for the eigenvalue to be negative, we have to impose that
\begin{equation}
 \pm\frac{(2\beta\kappa_1-\alpha \kappa_2)(\beta\kappa_2-2\alpha\kappa_3)^9}{(\beta^2 \kappa_1 - \alpha \beta \kappa_2+\alpha^2\kappa_3)} <0 \,.
\end{equation}
Close to the critical point, when $\bar{P}_\chi \to 0$, we see that $m_{33}\to -\infty$. The fact that the eigenvalue goes to infinity at the critical point is not relevant. The only relevant point for the stability is that it goes to $-\infty$ by imposing the above condition on the parameters of the theory. While one of the critical points will represent an attractor, the other one will necessarily be a repeller, since the two critical points only differ in the overall sign of $H$. 
Finally, the third value for the pressure $P^{(III)}_\chi$ substituted back into $H$ gives
\begin{equation}
H^{(III)}=\frac{\pm m}{\sqrt{6}(\beta^2\kappa_1-\alpha^2\kappa_3)} \sqrt{\mathcal{D}} \,,
\end{equation}
where the quantity under the square root stands for
\begin{eqnarray}
\mathcal{D}=(\frac1\beta (-10\alpha\beta^4\kappa_1^3-2\alpha^5\kappa_1\kappa_3^2-\beta^5\kappa_1^2(6\kappa_1+3\kappa_2+2\kappa_3)+2\alpha_3\beta^2\kappa_1(-3\kappa_2^2+2\kappa_1\kappa_3) \nonumber\\
+\alpha^4\beta(-6\kappa_1\kappa_3^2+(\kappa_2-2\kappa_3)(\kappa_2+\kappa_3)^2)+2\alpha^2\beta^3\kappa_1(6\kappa_1(\kappa_2+\kappa_3)+\kappa_3(3\kappa_2+2\kappa_3)))\,.
\end{eqnarray}
These are the two other critical points, that the system admits. Similarly, we analyze the linearized system around the two critical points characterised by $(H^{(III)},P^{(III)}_\chi,\Omega_\phi=0)$ and it turns out that one of them is again an attractor whereas the other one is a repeller. 

\begin{figure}[h!]
\includegraphics[width=8.1cm]{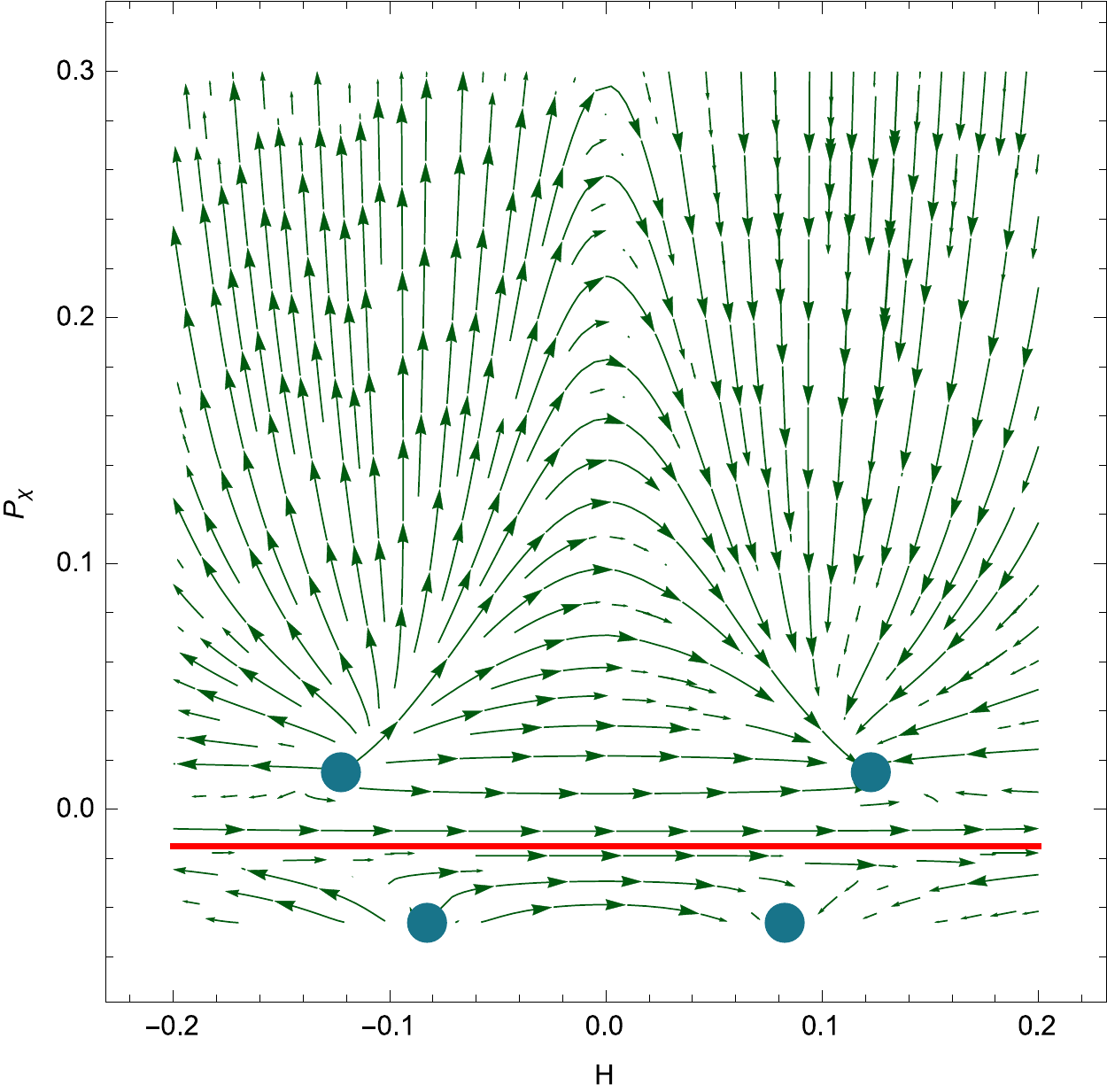}
\includegraphics[width=8.1cm]{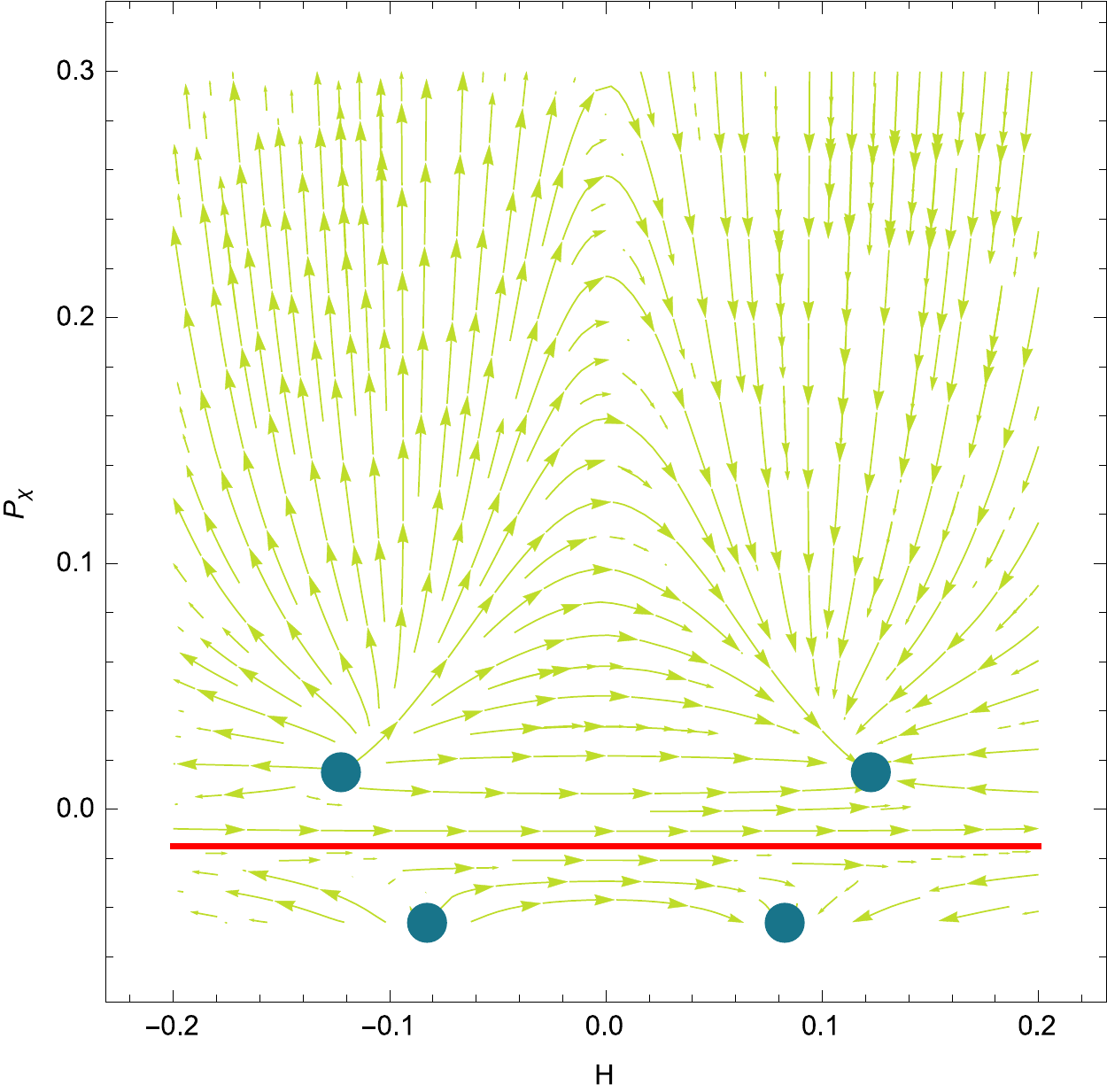}
\caption{This figure shows two examples of phase map portraits of the dynamical autonomous system for $\kappa_1=3$, $\kappa_2=-2$, $\kappa_3=-12$, $\alpha=1$ and $\beta=2$ with $\Omega_\phi=0.5$ in the left panel and $\Omega_\phi=0$ in the right panel. The parameters are chosen such that they fulfil the stability conditions given in the main text. The colour encodes $\dot{\Omega}_\phi$ being positive in the left panel and negative in the right panel respectively. The red line denotes the separatrix and the green points represent the critical points. One can see that two of them with positive $H$ are attractors whereas with negative $H$ are repellers. Very close to the separatrix, we see that the trajectories barely evolve in the pressure. One also observes immediately the interesting trajectories that start with a negative $H$ and evolve towards the de Sitter critical point with positive $H$. }
\label{phasemap}
\end{figure}

The behaviour of the dynamical autonomous system for a given choice of the parameters is illustrated in Fig. \ref{phasemap}. This corresponds to the $P_\chi-H$ plane where the variation of the standard matter field is encoded in the colour with positive and negative variation in the left and right panel respectively. One immediately recognises the four critical points where two pairs differ only in the sign for $H$. Another interesting feature is the attracting nature of the critical points with positive $H$. They constitute the stable de Sitter critical points. The presence of these de Sitter critical points shows that the model can be used successfully as an alternative to dark energy. Even in the presence of the standard matter fields, the system evolves towards these de Sitter critical points. The initial conditions will dictate if and which of the attractor de Sitter point will be achieved. 

\begin{figure}[h!]
\includegraphics[width=8.1cm]{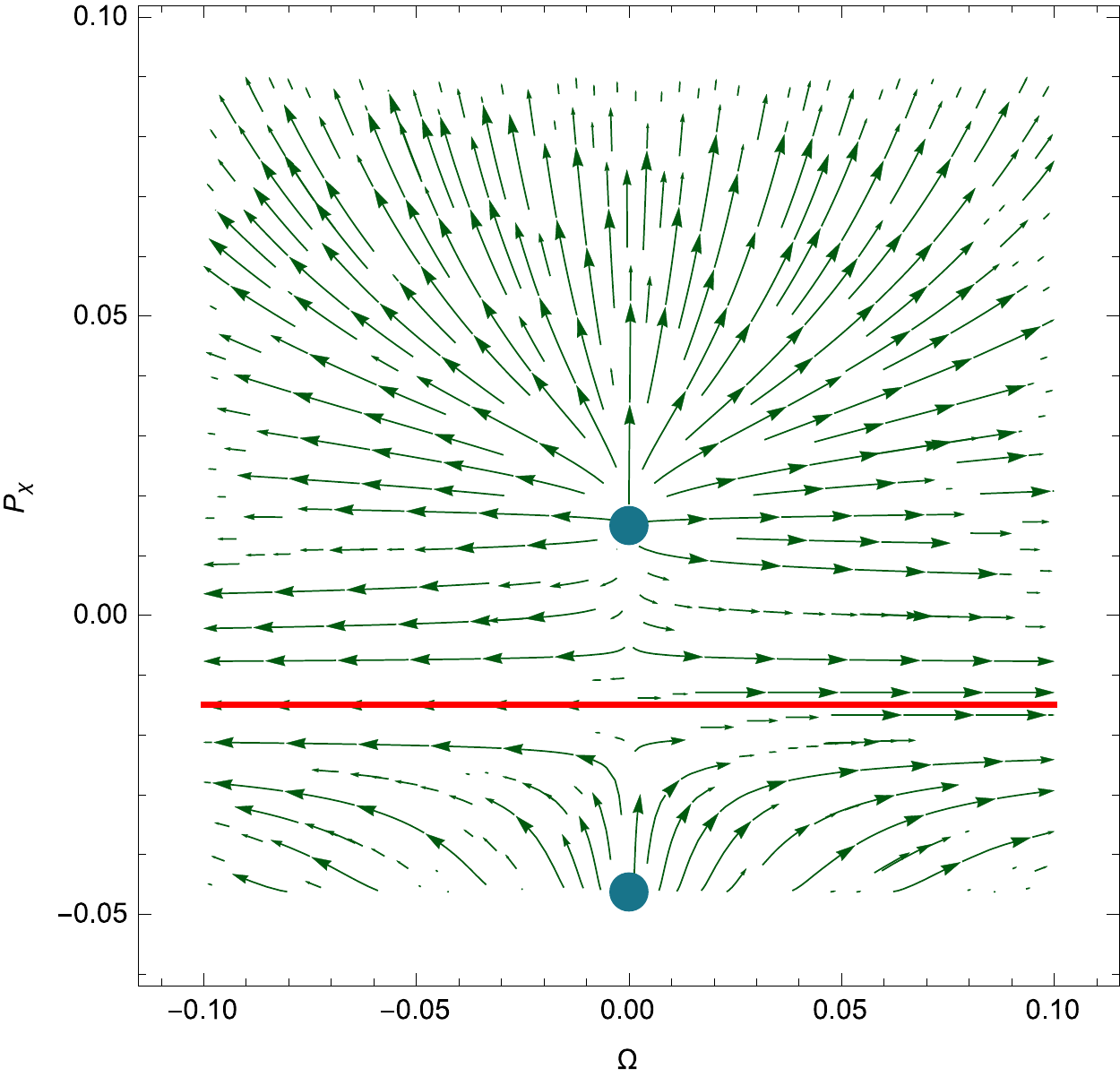}
\includegraphics[width=8.1cm]{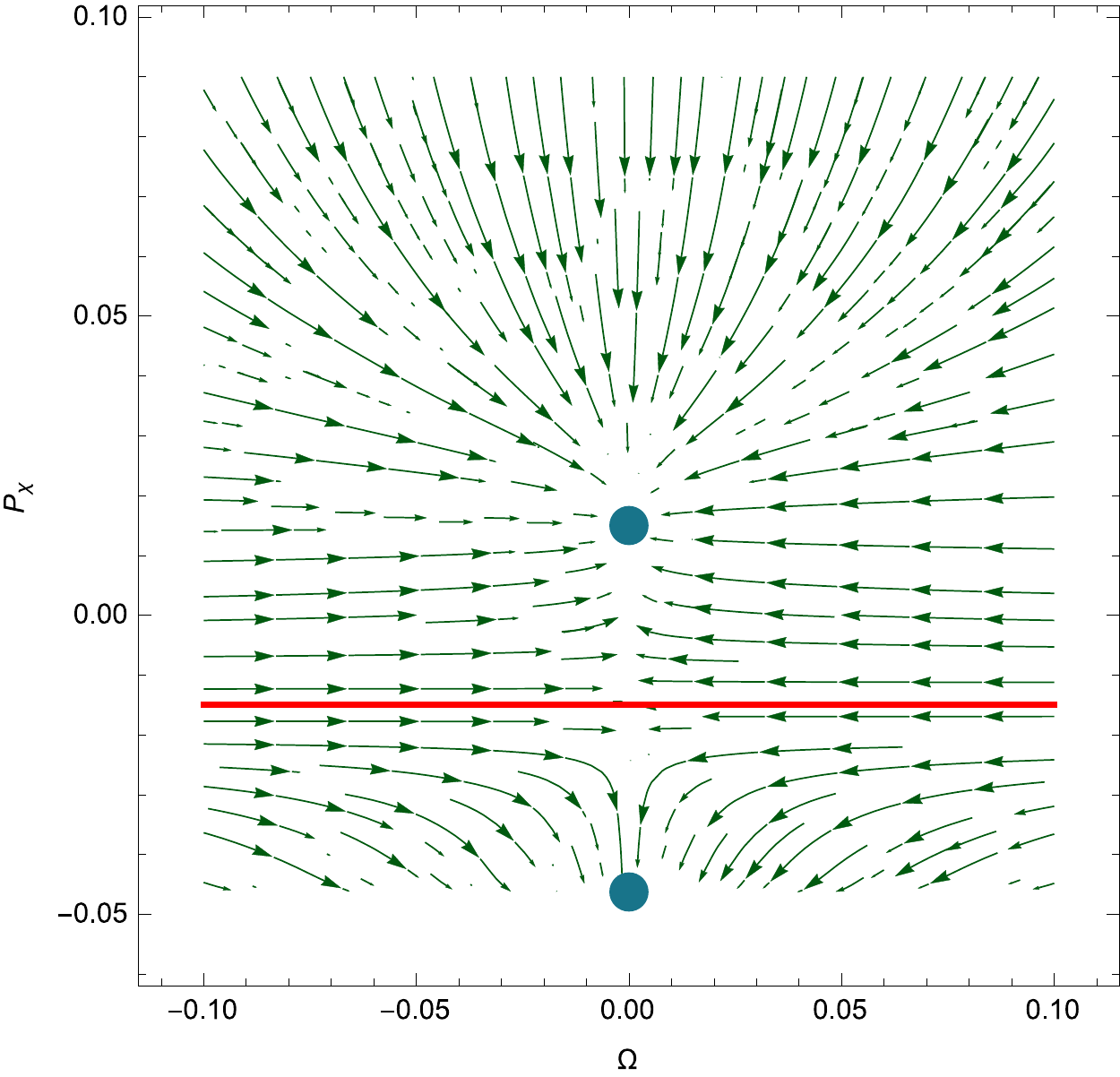}
\caption{Phase map portrait of the dynamical autonomous system as in Fig. \ref{phasemap}. The same parameters have been chosen. In order to plot the $\Omega_\phi-P_\chi$ plane, we have fixed the value for $H$ to be the value at the critical points $H^{(III)}$.  In the left panel it is the negative value for $H^{(III)}$ and in the right panel it is the positive value. The red line denotes again the separatrix and the green points are the critical points. We see the attractor nature of the critical point with positive $H$ in the right panel.}
\label{phasemapOmegay}
\end{figure}

As one can see, the two stable de Sitter attractors are separated by the presence of a separatrix at the value of the pressure given by $P_\chi=\frac{m^2\mpl^2\kappa_3}{\alpha\beta^3}$. Thus, if we start with an initial condition at the right corner in the lower part of the phase map with a pressure smaller than the value at the separatrix, we can never reach the de Sitter critical point above the separatrix in the upper part of the phase map. We see also in the phase map that there is no Minkowksi vacuum solution as an attractor solution for this particular choice of the parameters of the example. The Minkowski vacuum solution can never be an attractor solution in this model, and can be at most a saddle point.

Close to the separatrix one encounters many constant $P_\chi$ solutions. Also some trajectories between the critical points have mildly changing $P_\chi$. The trajectories going from the left critical point with negative $H$ to the right critical point with positive $H$ could give alternatives for early universe applications without singularities. The universe would start with a contracting phase and go through $H=0$ and end up in an expanding universe and reach the stable de Sitter critical point at some point. This can be clearly seen in Fig. \ref{phasemap}. 

\begin{figure}[h!]
\includegraphics[width=8.1cm]{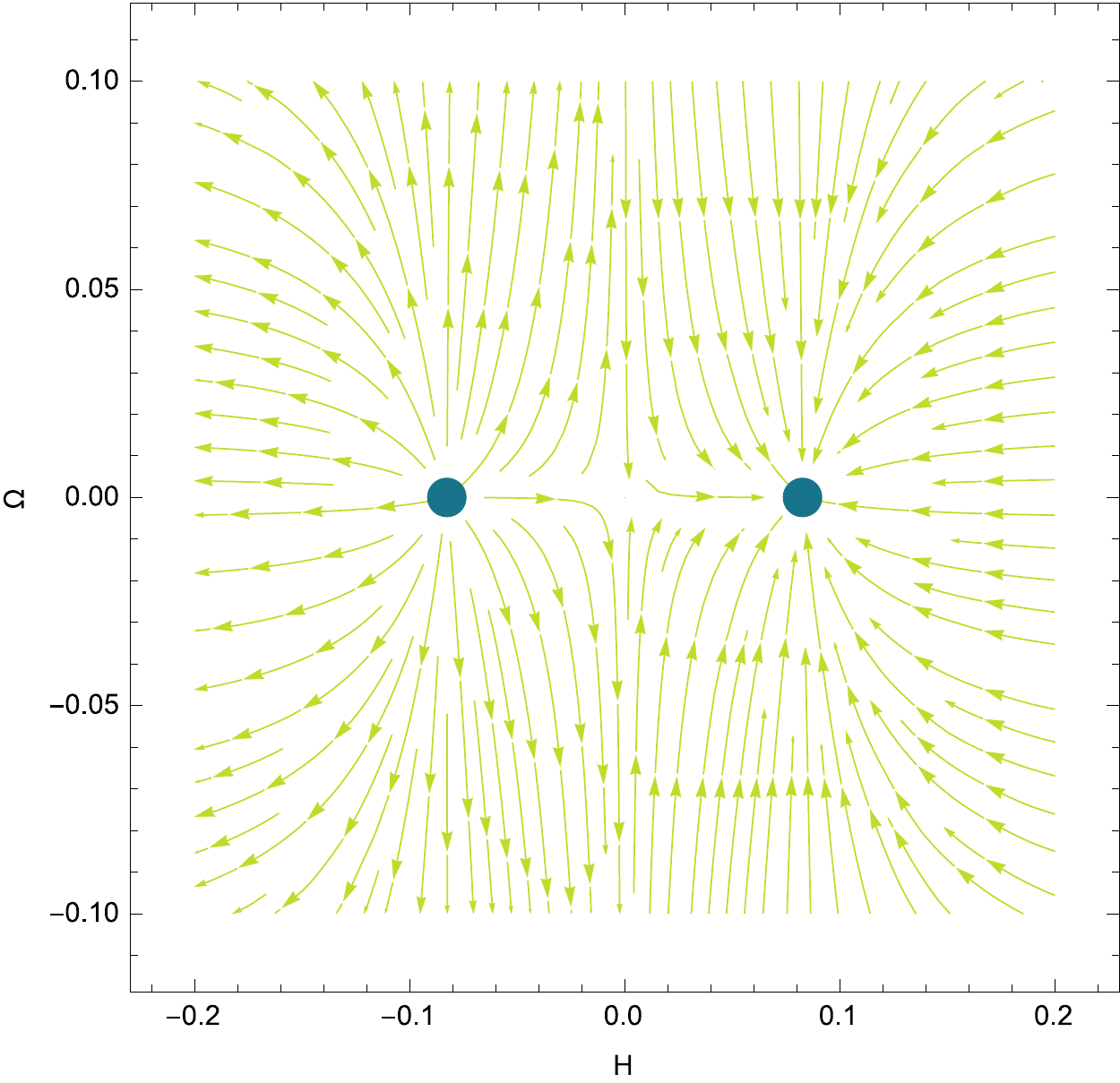}
\includegraphics[width=8.cm]{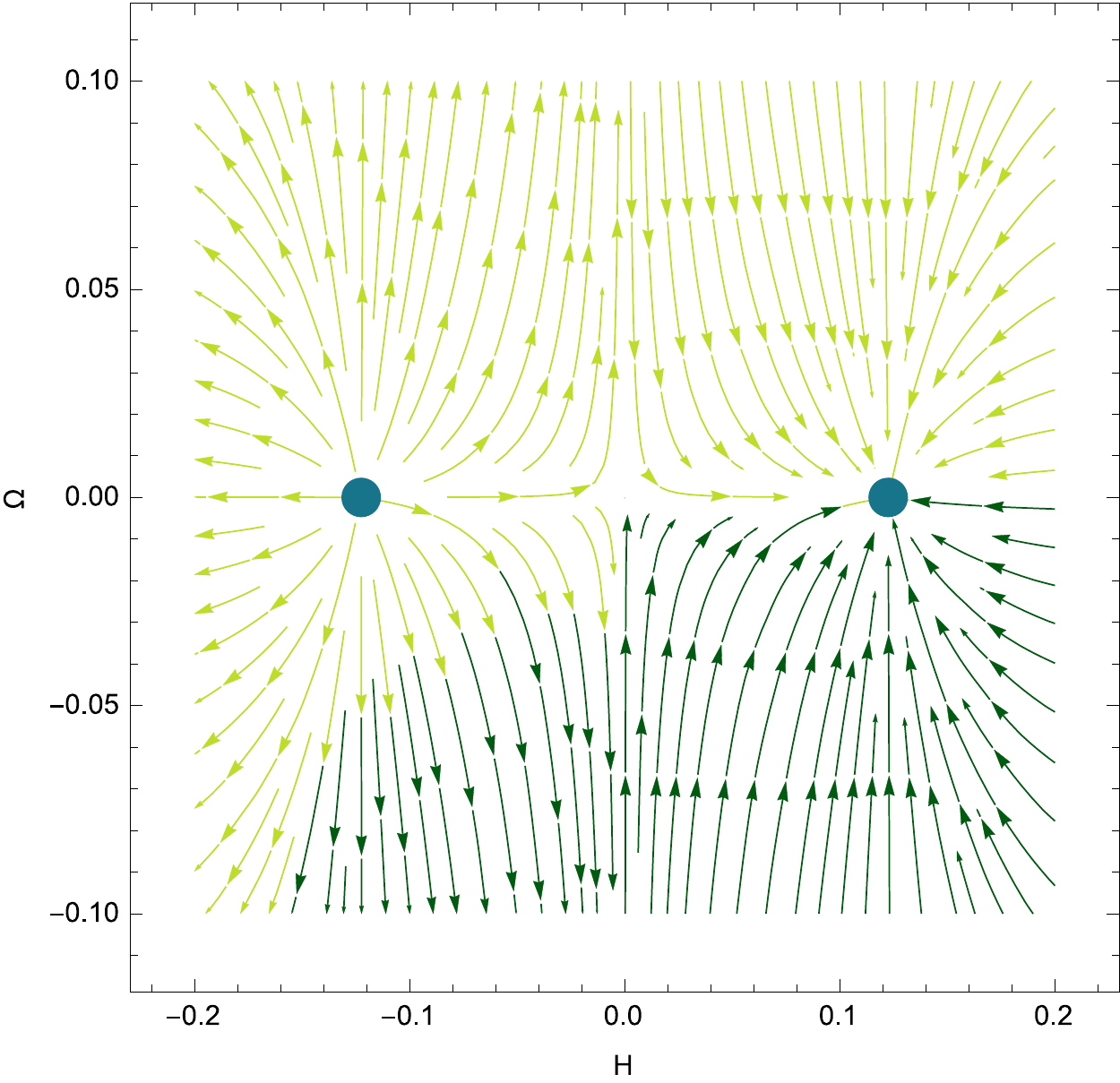}
\caption{For completeness we show the same phase map portrait as in Figs. \ref{phasemap} and \ref{phasemapOmegay} but in the $\Omega_\phi-H$ plane. This time we fix the value of $P_\chi$ to be $P_\chi^{(II)}$ in the left panel and $P_\chi^{(III)}$ in the right panel. The colour of the arrows encodes this time $\dot{P}_\chi$, the darker one being positive and lighter one being negative respectively. }
\label{phasemapOmegah}
\end{figure}

Additionally, the trajectories above the separatrix in the right upper panel represent (quasi) de Sitter solutions with mildly changing $H$ during the whole evolution until they come closer to the critical point. In these mentioned solutions $H$ remains nearly constant for some time and yields a period of quasi de Sitter expansion along these trajectories. The duration of this period depends on the parameters and the initial conditions. For the right expansion history as a dark energy model, we are more interested in the trajectories that start off at a given value for $H$ and decreases with time until the attractor de Sitter point is reached. In a similar way we also show the same phase map portrait of the same dynamical autonomous system in the $\Omega_\phi-P_\chi$ plane in Fig. \ref{phasemapOmegay}. Fig. \ref{phasemapOmegah} also gives the phase map portrait in the $\Omega_\phi-H$ plane with varying $P_\chi$ depicted by the colour.

 In Fig. \ref{numsolution} we plot the numerical evolution of one particular solution and one can clearly see the same behaviour as the analytic stability analysis resulted in. It is worth to emphasise that the presence of the standard matter field does not enlarge the number of critical points. All existing critical points have $\Omega_\phi=0$.

\begin{figure}[h!]
\includegraphics[width=8.cm]{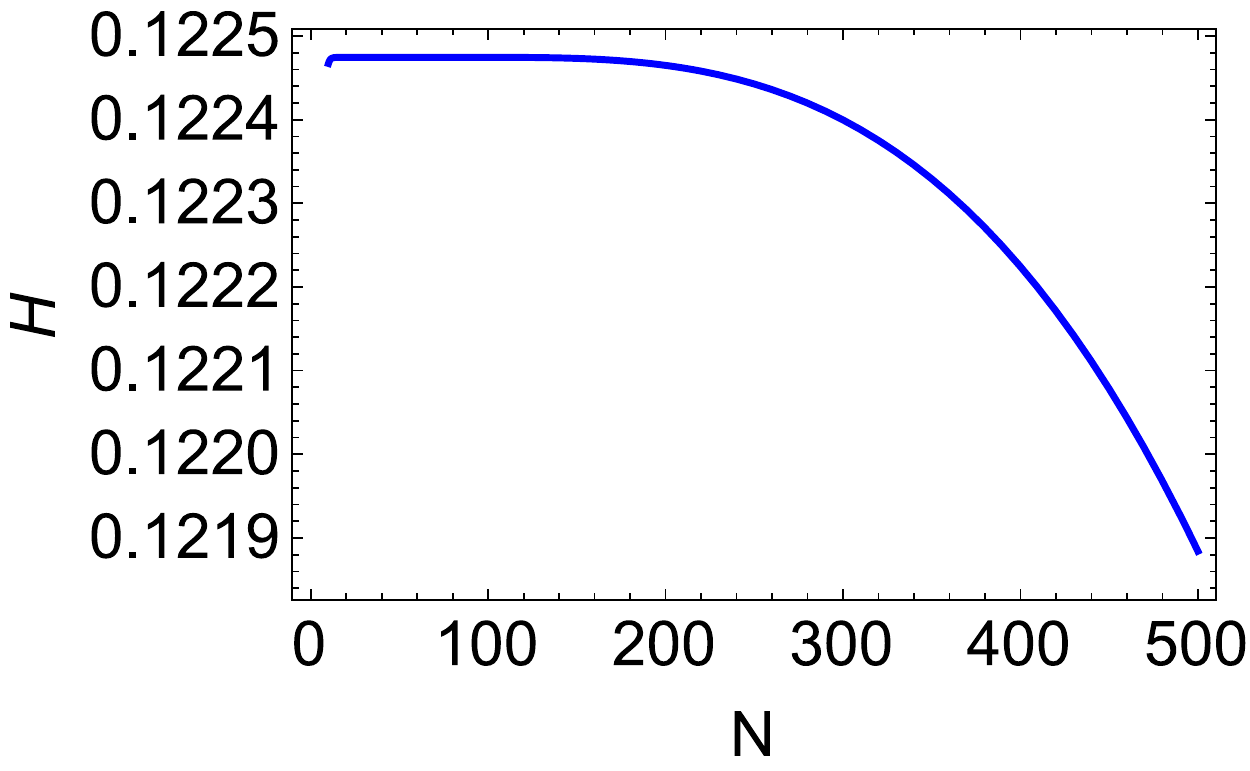}
\includegraphics[width=8.cm]{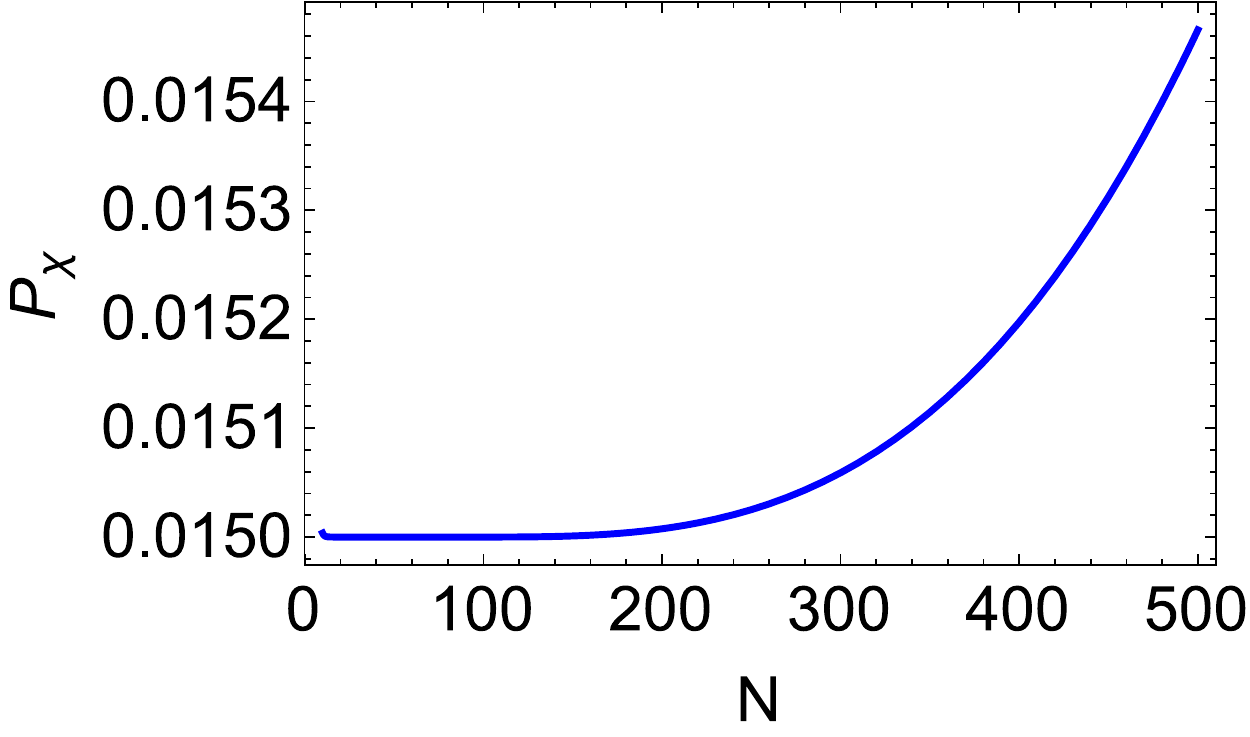}\\
\caption{This figure illustrates the numerical solution for $H$ in the left panel and $P_\chi$ in the right panel with the initial conditions $H_{ini}=0.1$ and $P^{ini}_\chi=0.3$ and $\Omega_\phi=0.5$. One sees the period of quasi de Sitter expansion in the evolution of $H$ lasting for approximately $200$ e-folds, that we were observing above the separatrix in the right upper panel of the phase map in Fig. \ref{phasemap}.}
\label{numsolution}
\end{figure}

\begin{figure}[h!]
\includegraphics[width=8.cm]{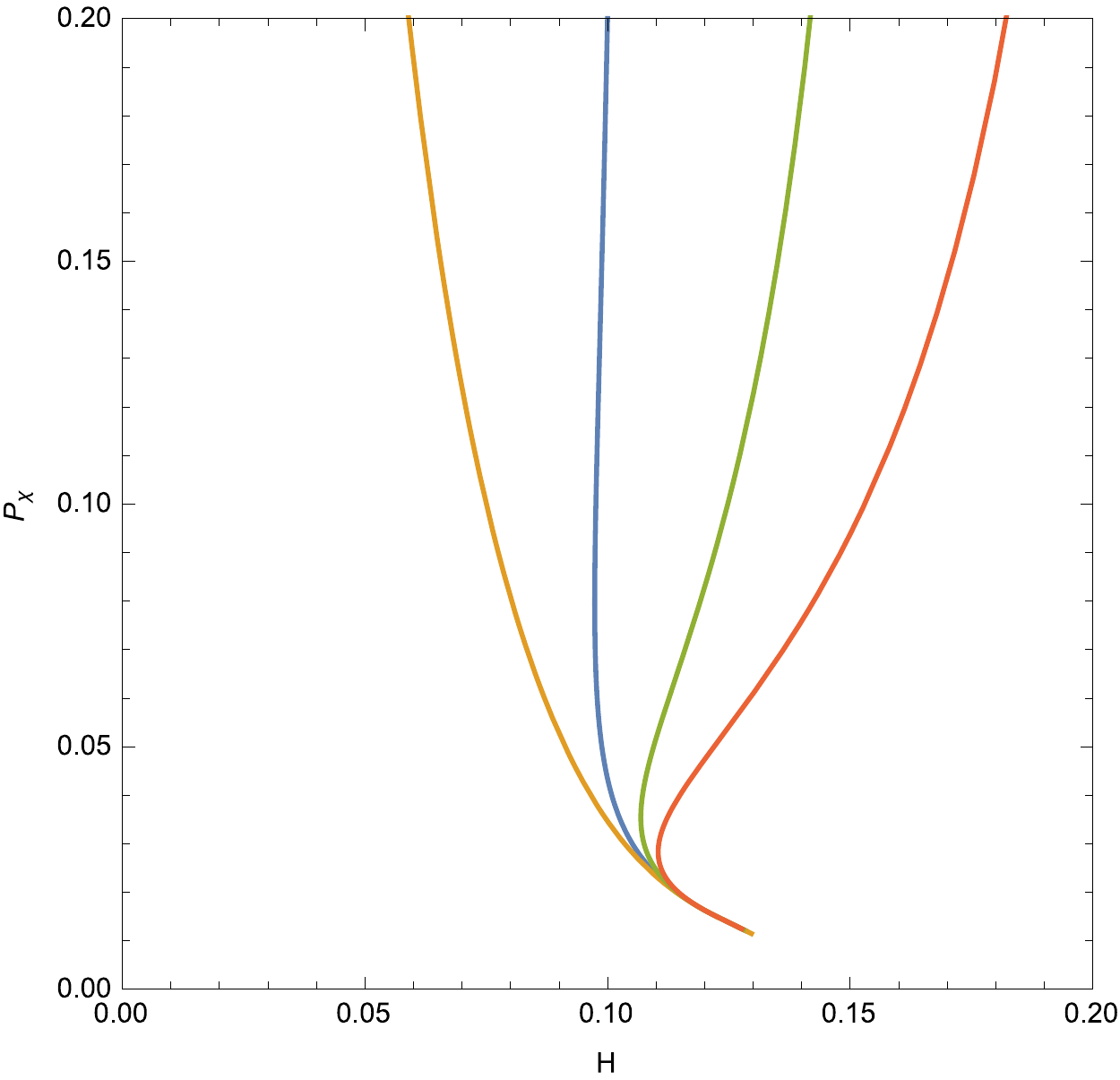}
\caption{In this figure we show the parametric plot of the numerically solved functions $H$ and $P_\chi$ for four different but close initial conditions. These indeed resemble the trajectories found above the upper separatrix in the right panel of the phase map.}
\label{numsolution}
\end{figure}


\section{The pure $\alpha_2$ case}\label{sec:alpha2}
In order to gain more intuition on the background equations of motion, we consider the special case $\alpha_3=\alpha_4=0$. This will allow us to write down explicitly the cumbersome expressions for some of the functions, that we have omitted in the previous section. Since the remaining parameter $\alpha_2$ is multiplied by an overall factor $m^2$, we can absorb $\alpha_2$ into $m$, so we can simply assume $\alpha_2=1$. Starting with the constraint equation \ref{eq:eqf}, we solve it for $A$
\begin{equation}
A=-\frac{\alpha^2\beta^2 P_\chi + \mpl(m^2\mpl+ \tilde{P}_\chi)}{\alpha\beta^3 P_\chi} \,,
\end{equation}
where we have introduced this time the shortcut notation $\tilde{P}_\chi=\pm\left(m^4\mpl^2+m^2\alpha\beta^2(2\alpha+3\beta)P_\chi\right)^{\frac12}$ for convenience. We again assume that the matter field that couples to the standard $g$ metric is pressureless, i.e. $P_\phi=0$ and the energy density can be rewritten in terms of the variable $\Omega_\phi=\rho_\phi/(6\mpl^2H^2)$. We can use the derivative of the Friedmann equation to solve for $\dot{H}$ and replace $A$ by the above expression. By doing that we obtain the first equation of the autonomous system in the form
\begin{eqnarray}
\frac{dH}{dN}= \frac{1}{2\alpha^2\beta^6P_\chi^2}(3\alpha^2\beta^4(m^2(\alpha+\beta)(\alpha+2\beta)+\beta^2H^2)P_\chi^2+2m^4\mpl^3(m^2\mpl+\tilde{P}_\chi) \nonumber\\
+m^2\mpl\alpha\beta^2(2\alpha+3\beta)P_\chi(3m^2\mpl+2\tilde{P}_\chi)) \,,
\end{eqnarray}
where we again used $dN=Hdt$. Similarly, we can bring the equations of motion for the matter fields in a similar form. For the standard matter field it reads
\begin{eqnarray}
\frac{d\Omega_\phi}{dN}= \frac{m^2\Omega_\phi}{\alpha^2\beta^6HP_\chi^2}(3\alpha^2\beta^4(\alpha+\beta)(\alpha+2\beta)P_\chi^2+2m^2\mpl^3(m^2\mpl+\tilde{P}_\chi) \nonumber\\
+\mpl\alpha\beta^2(2\alpha+3\beta)P_\chi(3m^2\mpl+2\tilde{P}_\chi)) \,,
\end{eqnarray}
whereas for the matter field living in the effective metric it becomes
\begin{eqnarray}\label{dPdN}
\frac{dP_\chi}{dN}= \frac{2H}{m^2\mpl\alpha\beta(2\alpha+3\beta)^2}(3m^2\mpl^2(m^2\mpl-\tilde{P}_\chi) \nonumber\\
+\alpha\beta(2\alpha+3\beta)P_\chi(3m^2\mpl\beta+\alpha\tilde{P}_\chi)) \,.
\end{eqnarray}
We now diagnose the critical points for this specific choice of the parameters. We can solve from the vanishing of $\frac{dH}{dN}$ the corresponding value for H and plug this back into the vanishing of $\frac{d\Omega_\phi}{dN}$. By doing that we observe the familiar result from the previous section that, at the critical points, we have $\Omega_\phi=0$. Taking the values for $H$ and $\Omega_\phi$ at the critical points and plugging them into the vanishing of $\frac{dP_\chi}{dN}$ gives the critical points for $P_\chi$. The first value for the pressure that makes $\frac{dP_\chi}{dN}$ vanish is uninteresting for this choice of parameters with pure $\alpha_2$ term. It does not represent any critical point, and any separatrix either. Even though the expression for $H^{(I)}$ becomes complex infinity for $P^{(I)}_\chi=0$, asymptotically nearing the point $P_\chi=P^{(I)}_\chi+\epsilon$ does not yield anything unusual which becomes clear from the autonomous system near that point
\begin{eqnarray}
\frac{dH}{dN}&\approx&\frac38(m^2-4H^2)+\mathcal{O}(\epsilon) \nonumber\\
\frac{d\Omega_\phi}{dN}&\approx&-\frac3{4H}m^2\Omega_\phi+\mathcal{O}(\epsilon)   \nonumber\\
\frac{dP_\chi}{dN}&\approx&H\epsilon+\mathcal{O}(\epsilon^{2}) \,.
\end{eqnarray}
Thus, for the pure $\alpha_2$ case that we are considering in this section, we will not encounter any separatrix in the phase map. As next we shall consider the remaining two values for $P_\chi$ that makes $\frac{dP_\chi}{dN}$ in equation \ref{dPdN} vanish
\begin{equation}
P^{(II)}_\chi=\frac{3m^2\mpl^2}{\alpha^3\beta}\,,\qquad P^{(III)}_\chi=-\frac{m^2\mpl^2}{\alpha\beta^2(2\alpha+3\beta)} \,.
\end{equation}
The expression for $H$ at these two critical points depends on $P_\chi$ at the critical points, hence we can plug the expression for $P_\chi$ into $H$ to obtain
\begin{equation}
H^{(II)}=\frac{\pm m}{3\sqrt{3}\beta^2}\mathcal{J} \,,\qquad H^{(III)}=\frac{\pm m}{\beta}\sqrt{\frac{\alpha^2}{3}+\alpha\beta+\beta^2} \,,
\end{equation}
with the shortcut notation $\mathcal{J}=\left(-4\alpha^4-36\alpha^3\beta-108\alpha^2\beta^2-135\alpha\beta^3-54\beta^4\right)^{\frac12}$ introduced for convenience. Next we study the stability around the critical points. For this purpose, we compute the linearized system for each critical point 
\begin{eqnarray}
\frac{d}{dt}\left(\begin{array}{c}
\delta H \\ \delta  \Omega \\ \delta P_\chi \end{array}\right)  &=& M_{II,III} \left(
\begin{array}{c}
\delta H \\\delta  \Omega \\ \delta P_\chi\end{array}\right) 
\end{eqnarray}
with the corresponding matrices expressed as
\begin{eqnarray}
M_{II} =
\begin{pmatrix}
-3&0&\frac{\alpha^4(2\alpha+3\beta)^3}{6\sqrt{3}m\mpl^2\beta\mathcal{J}} \\
18&-3&\frac1{m^{2}} \\
0&0&5-\frac{8\alpha}{2\alpha+3\beta} 
 \end{pmatrix} \qquad \text{and} \qquad M_{III} =
 \begin{pmatrix}
-3&0&m^{III}_{13} \\
6&-3&m^{III}_{23}  \\
0&0&m^{III}_{33}  
 \end{pmatrix}
\end{eqnarray}
The eigenvalues of $M_{II}$ correspond to $\lambda^{(II)}_1=-3$, $\lambda^{(II)}_2=-3$ and $\lambda^{(II)}_3=5-\frac{8\alpha}{2\alpha+3\beta}$. In order for the critical point to be stable, we have to impose $\frac{8\alpha}{2\alpha+3\beta}>5$. Similarly the eigenvalues of $M_{III}$ are $\lambda^{(III)}_1=-3$, $\lambda^{(III)}_2=-3$ and $\lambda^{(III)}_3=m^{III}_{33} $. Note that $m^{III}_{33} $ diverges at the critical point $P^{(III)}_\chi$ and is of the form $m^{III}_{33}\sim -m^2\mpl(\alpha+3\beta)/(2\alpha+3\beta) \times 1/(P_\chi-P^{(III)}_\chi)$. But as long as $-m^2\mpl(\alpha+3\beta)/(2\alpha+3\beta)$ is kept negative, the critical point is stable. 

\begin{figure}[h!]
\includegraphics[width=8.cm]{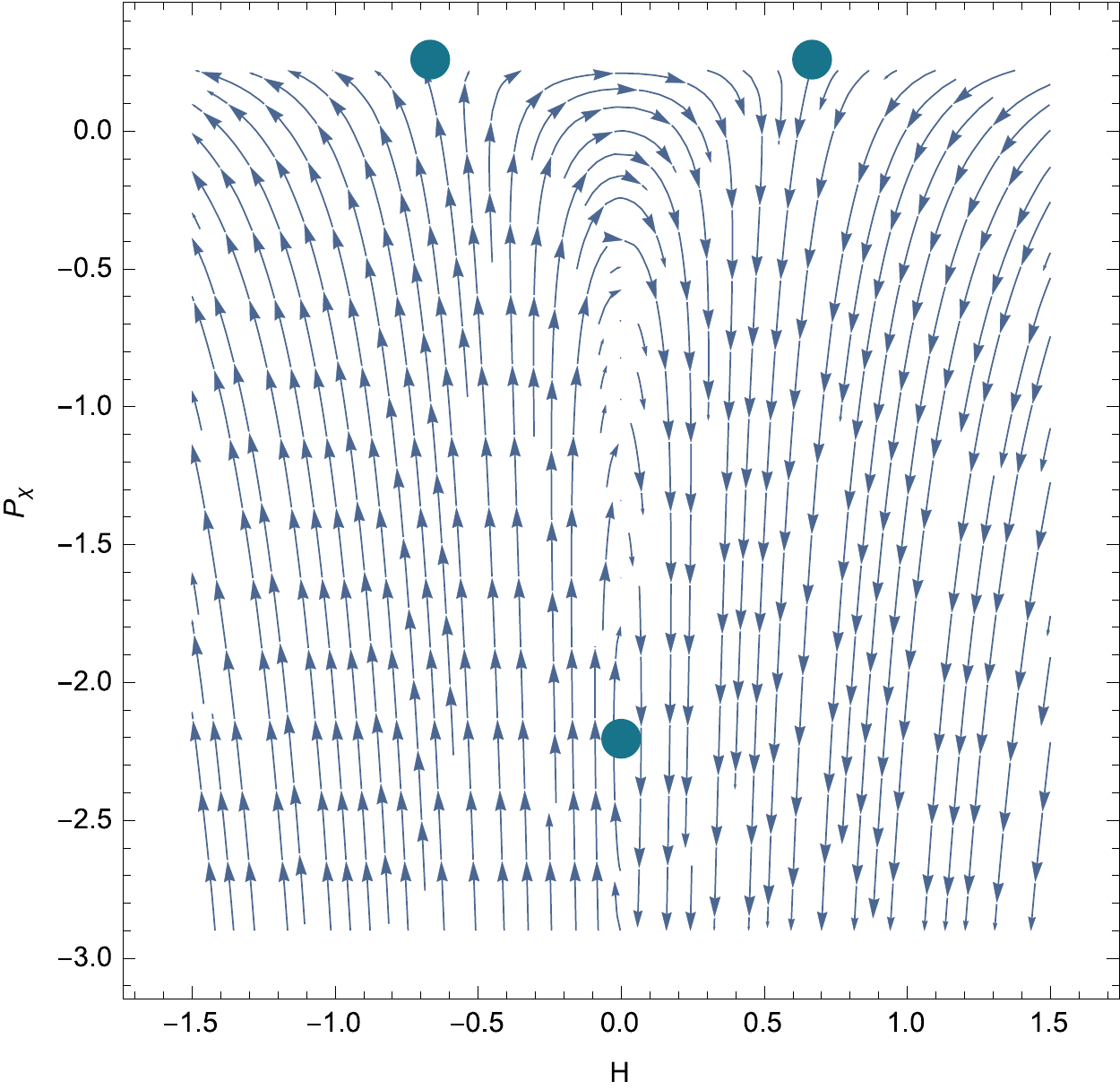}
\caption{In this figure we show one concrete example of phase map portrait of the dynamical autonomous system for the pure $\alpha_2$ case with $\kappa_1=3$, $\kappa_2=-2$, $\kappa_3=0$, $\alpha=1$ and $\beta=-1.36$ and with $\Omega_\phi=0$. For the special case of this choice of the parameters there is no separatrix related to $P^{(I)}_\chi$. However, one pair of the critical points become an attracting ( with negative $H$) or repelling barrier (with positive $H$) and hence the pressure of the matter of the dark sector can not be larger than $P^{(III)}_\chi$ for this particular choice of parameters. The other pair of $H$ converge into a single critical point with $H=0$. Thus, in this example Minkowksi vacuum solution is a critical point that is stable in one direction but unstable in the other, corresponding to a saddle point. }
\label{phasemap2}
\end{figure}

We plot a simple example for the $\alpha_2$ case in Fig. \ref{phasemap2}. The first immediate observation is the disappearance of the separatrix in the phase map coming from $P^{(I)}_\chi$, in agreement with our earlier analysis where we considered a small departure of the vanishing $P^{(I)}_\chi$ and took the corresponding limit. Another important observation is the merging of the two critical points into a single point with $H=0$. This is the Minkowski vacuum solution. Since the repeller and attractor critical points merge together, this solution can not be stable. In fact, it corresponds to a saddle point. To be more precise, it is stable in one direction and unstable in the other direction since one of the eigenvalues becomes positive. The third observation is the fact that the two critical points at $P^{(III)}_\chi$ become a barrier in the sense that the values of $P_\chi$ can not be larger than the critical value $P_\chi<P^{(III)}_\chi$ since the equations of the autonomous system become complex.

We have seen that in the general case as well as in this simple case of the parameter space there are stable de Sitter critical points. The parameters should be chosen such that the background evolution will be very similar to the standard model of cosmology with a cosmological constant. The presence of de Sitter critical points is crucial for the right phenomenology of the late-time universe. This should be compared to observations that constrain the background dynamics in order to further restrict the allowed parameter space. In the next section, we study the stability of the dynamical background equations. 


\section{Stability of the perturbations}\label{sec:perturbations}
We would like to study the stability of the perturbations around the homogeneous and isotropic background that we considered in the previous section. We will impose the absence of ghost and gradient instabilities of the tensor, vector and scalar perturbations. For this purpose, let use perturb the dynamical metric $g_{\mu\nu}$ in the following way
\begin{eqnarray}\label{perturbed_metric}
\delta g_{00} &=& -2\,N^2\,\Phi\,,\nonumber\\
\delta g_{0i} &=& N\,a\,\left(\partial_i B+B_i\right)\,,\nonumber\\
\delta g_{ij} &=& a^2 \left[2\,\delta_{ij}\psi +\left(\partial_i\partial_j-\frac{\delta_{ij}}{3}\partial^k\partial_k\right)E+\partial_{(i}E_{j)}+h_{ij}\right]\,.
\end{eqnarray}
This decomposition is the irreducible representation of the spatial rotations. In contrast to the background quantities,  all the perturbations are functions of time and space. We also note that $\delta^{ij}h_{ij} = \partial^ih_{ij} = \partial^i E_i = \partial^i B_i=0$ and that we will keep the St\"uckelberg fields purely background. This of course fix the gauge freedom completely. For the matter fields that couple to the effective metric and the standard matter field that couples to the dynamical metric, we had assumed a general scalar field. Therefore, they will only contribute to the scalar perturbations. For the compatibility with the background symmetry, we perturb the scalar fields $\chi$ and $\phi$ as follows
\begin{equation}
\chi=\chi_0(t)+\mpl \delta\chi\,, \qquad \text{and} \qquad
\phi=\phi_0(t)+\mpl \delta\phi\,.
\end{equation}
Going through the number of degrees of freedom in the action \ref{action_MG_effcoupl} one encounters na\"ively counted tvelve
degrees of freedom (dof). Out of these two are represented by the traceless symmetric spatial tensor fields ($h_{ij}$), four by the divergence-free spatial vector fields ($B_i$, $E_i$) and the remaining six dof by the scalars ($\Phi$, $B$, $\psi$, $E$, $\delta\chi$, $\delta\phi$). Not all of these dof propagate. In fact, one immediately observes that two of the scalar fields ($\Phi$, $B$) and two of the vector fields ($B_i$) are non-dynamical and one can integrate them out. The construction also guarantees that the Boulware-Deser ghost is absent and hence we will be able to integrate out one more combination. Finally, the remaining physical dof are two tensors, two vectors and three scalars dof. They correspond to nothing else than the five polarizations of the massive spin--2 field and the two matter field $\chi$ and $\phi$. Our homogeneous background allows us to decompose all the perturbations in Fourier modes with respect to the spatial coordinates, which will be used in the remaining of the section.

We will first compute  the quadratic action in the tensor perturbations. For this purpose, we substitute our ansatz in equation \ref{perturbed_metric} for the metric perturbations into the Lagrangian in equation \ref{action_MG_effcoupl}, decompose the tensor field in Fourier modes and make an extensive use of the background equations. The action for the tensor modes can then be expressed as
\begin{equation}
S^{(2)}_{\rm tensor} = \frac{\mpl^2}{8}\int d^3k\,dt\,N\,a^3\,\left[\frac{1}{N^2}\dot{h}_{ij,\vec{k}}^\star \dot{h}^{ij}_{\vec{k}}-\left(\frac{k^2}{a^2}+m_{T}^2\right)h_{ij,\vec{k}}^\star h^{ij}_{\vec{k}} \right]\,,
\end{equation}
with the mass term of the tensor perturbations represented by
\begin{equation}
m_{T}^2\equiv m^2(r-1)A(J(A-1)^2(\alpha+2(\alpha+\beta)A+\beta A^2)-(\alpha+\beta A)(Q+A^2\rho_m))(A-1)^{-3}(\alpha+\beta A)^{-1} \,.
\label{eq:MGW2}
\end{equation}
The tensor perturbations are the same as in \cite{Gumrukcuoglu:2014xba} since the additional matter field does not contribute to the tensor perturbations. They already have the right sign for the kinetic term. So there are no associated ghost instabilities of the tensor modes. The same is true for the gradient instabilities either. In order to avoid  tachyonic instability, one has to impose $m_{T}^2>0$. Note, that the time-scale of the instability is associated to the inverse graviton mass.

Next, we study the stability of the vector perturbations. Similarly to the tensor modes, we first decompose the vector modes into their Fourier modes and expand our action\ref{action_MG_effcoupl}  to second order in the vector perturbations. The first thing that one observes is that the vector modes $B_i$ are actually non-dynamical and we can use their equations of motion to express them in terms of the vector modes $E_i$. After integrating out the $B_i$ modes, the quadratic action in the vector modes simply becomes
\begin{equation}
S^{(2)}_{\rm vector} = \frac{\mpl^2}{16}\int d^3k\,dt\,k^2a^3 \left[m_V^2
\dot{E}_{i,\vec{k}}^\star\dot{E}^i_{\vec{k}}-m_{T}^2 E_{i,\vec{k}}^\star E^i_{\vec{k}} 
\right]\,.
\end{equation}
The function in front of the kinetic term $m_V^2 \equiv n_1/d_1$ has the numerator and denominator given by
\begin{eqnarray}
n_1&=&2a^2A(\alpha(2+r)+\beta(1+2r)A)(\tilde{\mathcal{F}}N^2-2\beta\dot{\phi}^2\partial_{X_\phi}\tilde{P}(X_\phi,\phi) ) \nonumber\\
d_1&=&N^2(\tilde{r}\alpha^2+A(\tilde{r}\beta(2\alpha+\beta A)+2\tilde{\mathcal{F}}a^2(\alpha(2+r)+\beta(1+2r)A)))\nonumber\\
&-&4\beta a^2A(\alpha(2+r)+\beta(1+2r)A)\dot{\phi}^2\partial_{X_\phi}\tilde{P}(X_\phi,\phi) 
\end{eqnarray}
where we introduced the short-cut notation $\tilde{r}=k^2\mpl^2(1+r)^2$ and $\tilde{\mathcal{F}}=3\mpl^2\beta H^2+\beta \tilde{P}(X_\phi,\phi)+m^2\mpl^2(J(\alpha+\beta A)-\beta \rho_m)$. We recover the result obtained in \cite{Gumrukcuoglu:2014xba} for vanishing $\phi$, i.e. in the absence of the additional matter field. 

Next, we study the stability conditions of the scalar perturbations. As mentioned above, the difference to the study in \cite{Gumrukcuoglu:2014xba} will come in this sector due to the presence of the standard matter field in form of a general scalar field. As before, we compute the action quadratic in scalar perturbations and introduce their Fourier modes. Out of the six scalar modes $\Phi$, $B$, $\psi$, $E$, $\delta\chi$ and $\delta\phi$, the two scalar field $\Phi$ and $B$ are non-dynamical. 
This is already visible in the corresponding kinetic matrix, which has two vanishing eigenvalues, imposing three constraint equations
\begin{eqnarray}
\mathcal{K}_{\psi, \delta \chi, \delta\phi, E, B, \Phi}=
\begin{pmatrix}
-6&0&0&0 &0&0\\
0&\mathcal{K}_{ \delta\chi  \delta\chi} &0&0 &0&0\\
0&0&\mathcal{K}_{ \delta\phi  \delta\phi} &0 &0&0\\
0&0&0&k^4/6&0&0 \\
0&0&0&0&0&0 \\
0&0&0&0&0&0
 \end{pmatrix}
\end{eqnarray}
with $\mathcal{K}_{ \delta\phi  \delta\phi}=2\partial_{X_\phi}\tilde{P}(\phi, X_\phi)+4\dot{\phi}^2\partial^2_{X_\phi}\tilde{P}(\phi, X_\phi)/N^2$ and $\mathcal{K}_{ \delta\chi  \delta\chi}$ given by
\begin{equation}
\mathcal{K}_{ \delta\chi  \delta\chi}=\frac{N^2(\alpha+\beta rA)^4(N^2\tilde{\mathcal{F}}-2\beta\dot{\phi}^2\partial^2_{X_\phi}\tilde{P}(\phi, X_\phi))}{2\alpha\beta(\alpha+\beta A)^3\dot{\chi}^2}
+2\dot{\chi}^2\partial^2_{X_\chi}P(\chi, X_\chi)
\end{equation}
We can use the equations of motion for $\Phi$ and $B$ to integrate them out. The variation with respect to $B$ gives
\begin{eqnarray}
 \nonumber\\
 3\mpl\delta\chi_{\vec{k}}(\alpha+\beta r A)(\alpha+r(\alpha+2\beta A))(\tilde{\mathcal{F}}N^2-2\beta\dot{\phi}^2\partial_{X_\phi}\tilde{P}(\phi,X_\phi))(\beta(1+r)(\alpha +\beta A)^2\dot{\chi})^{-1}   \nonumber\\
-\frac{3n_1B_{\vec{k}}}{2aN(1+r)^2(\alpha+\beta A)^2}-6\mpl^2HN\Phi_{\vec{k}}-k^2\mpl^2\dot{E}_{\vec{k}}-6\mpl^2\dot{\psi}_{\vec{k}} -6\mpl \delta\phi_{\vec{k}}\dot{\phi} \partial_{X_\phi}\tilde{P}(\phi, X_\phi) =0  \nonumber
\end{eqnarray}
and we can solve it for $B$. Similarly we vary the quadratic action with respect to $\Phi$ and solve it for $\Phi$.
After integrating out the scalar modes $B$ and $\Phi$ the remaining action depends only on the four scalar modes $\psi$, $E$, $\delta\chi$ and $\delta\phi$. 
The kinetic matrix of these four scalar fields has a vanishing determinant, signalling that there is still a remaining constraint equation and we can integrate out one more non-propating degree of freedom, namely the Boulware-Deser mode. In fact, this becomes manifest after performing the right field redefinition. For this purpose we first compute the eigenvectors $v_1$, $v_2$, $v_3$ and $v_4$ of the kinetic matrix $\mathcal{K}_{\psi, \delta \chi, \delta\phi, E}$ and take the transpose of it
\begin{equation}
P=\{{v_1, v_2, v_3, v_4  \}}^\top \,.
\end{equation}
The matrix $P$ takes the basis $\psi, E, \delta \chi, \delta\phi$ and brings it into a basis, in which the Boulware-Deser mode $\psi$ becomes non-dynamical and can be easily integrated out using its equation of motion. The new variables are defined as
\begin{equation}
\left(
\begin{array}{cc}
\pi_{1,\vec{k}}  \\
\pi_{2,\vec{k}} \\
\pi_{3,\vec{k}} \\
\pi_{4,\vec{k}}
\end{array}
\right) = P^{-1}\left(
\begin{array}{cc}
\psi_{\vec{k}}  \\
E_{\vec{k}} \\
\chi_{\vec{k}} \\
\phi_{\vec{k}}
\end{array}
\right) 
\end{equation}
In term of these new field variables, we can integrate out one mode using its equation of motion. The remaining quadratic action depends only on the three propagating modes
\begin{equation}
S^{(2)}_{\rm scalar}= \frac{\mpl^2}{2}\int d^3k \,dt\,a^3 \,\left(\dot{\Pi}^\dagger\,\hat{K}\,\dot{\Pi} + \dot{\Pi}^\dagger\,\hat{{\cal N}}\,\Pi- \Pi^\dagger\,\hat{{\cal N}}\,\dot{\Pi}-\Pi^\dagger\,\hat{M} \,\Pi\right)\,,
\end{equation}
where $\Pi$ denotes $\Pi=\{ \pi_{2,\vec{k}}, \pi_{3,\vec{k}}, \pi_{4,\vec{k}}\}$and $\hat{K}$, $\hat{M}$ and $\hat{{\cal N}}$ are $3\times3$ real, time-dependent matrices. 
Their exact form are very cumbersome but we quote here their leading terms in the subhorizon limit. The kinetic matrix in this limit corresponds to
\begin{eqnarray}
\hat{K}\sim
\begin{pmatrix}
\frac{(\rho_\chi+P_\chi)}{2c_\chi^2X_\chi}\frac{ a_\eff^3/N_\eff}{a^3/N}&0&0\\
0&\frac{(\rho_\phi+P_\phi)}{2c_\phi^2X_\phi} &0\\
0&0&\frac{(\rho_\chi+P_\chi)\alpha\beta a_\eff^3 N A}{\mpl^2(a_\eff N(r+1)-N_\eff a)}
 \end{pmatrix} +{\cal O}(k^{-2})\,,
\end{eqnarray}
while the potential matrix has the following non-vanishing leading contributions
\begin{eqnarray}
\hat{M}_{11} &\sim&\frac{N_\eff^3a_\eff}{a^3N\dot{\chi}}\left( 1-\frac{a}{N_\eff}\frac{\alpha\beta AN^2(r-1)^2}{(a_\eff N(r+1)-aN_\eff)} \right)(\rho_\chi+P_\chi) k^2  +{\cal O}(k^0)\,,\nonumber\\
\hat{M}_{22} &\sim& \frac{(\rho_\phi+P_\phi)}{2\,a^2\,X_\phi}\,k^2+{\cal O}(k^0)\,,\nonumber\\
\hat{M}_{33} &\sim& {\cal O}(k^0)\,.
\end{eqnarray}
Finally, the mixing matrix with one derivative has the following non-vanishing component at order ${\cal O}(k)$
\begin{eqnarray}
\hat{N}_{13}\sim\frac{\alpha\beta Aa_\eff^2N_\eff N (r-1)(\rho_\chi+P_\chi) }{2\mpl a(-aN_\eff+(r+1)a_\eff N)\dot{\chi}}k+{\cal O}(k^0)\,.
\end{eqnarray}
In order to avoid ghost instability, we have to impose that the diagonal components of the above kinetic matrix are positive. The first diagonal component has the right sign if we impose $(\rho_\chi+P_\chi)>0$ whereas for the second one we have to impose $(\rho_\phi+P_\phi)>0$. Finally, for the third component to be positive we have to require that $(a_\eff N(r+1)-N_\eff a)>0$ together with $\alpha\beta>0$.

\section{Conclusions}
\label{sec:conclusion}

In this work, we have studied the cosmological implications of doubly coupled matter fields in the framework of massive gravity. This model does not only circumvent the no-go result for the existence of exact FLRW solutions in massive gravity, but also offers rich phenomenology. For this purpose, we have assumed that the doubly coupled matter field is a constituent of the dark sector. Furthermore, we have assumed that the standard matter field still couples only to the dynamical metric. For the general analysis of the cosmological solutions, we have performed dynamical system analysis. After bringing the background equations into the form of an autonomous system, we have investigated in detail the existence of critical points of the cosmological equations and their stability. 
We have seen that the system admits two pairs of critical points that differ only by an overall sign in the value of the Hubble rate $H$. While one of the pair corresponds to an attractor critical point, the other one necessarily represents a repeller. Thus, the system admits stable de Sitter critical points. All the critical points of the system were characterised by $\Omega_\phi=0$, meaning that even if the amount of matter was initially dominant, the system transits from a matter dominated universe to an accelerated phase. This is so because one of the critical point is always an attractor with $\Omega_\phi=0$. Thus the framework of massive gravity with doubly coupled matter field can play the role of dark energy. We have also shown the existence of a separatrix, whose presence depends strongly on the choice of the parameters. The separatrix separates the two pairs of critical points from each other. Furthermore, we have studied the stability of tensor, vector and scalar perturbations on top of FLRW background and worked out the conditions that have to be satisfied in order to avoid ghost and gradient instabilities. These results show that massive gravity in the presence of the effective composite metric can provide stable dark energy framework. Even if an attractor de Sitter critical point exists, this does not necessarily mean that the model will guarantee a good fit to observations. The constraints coming from the background observations will be considered in a future work as well as the consequences for the observations coming from the perturbations.

\acknowledgments

We would like to thank J. Beltran Jimenez, R. Brandenberger, T. Kacprzak and S. Seehars for useful discussions.
LH acknowledges financial support from Dr. Max R\"ossler, the Walter Haefner Foundation and the ETH Zurich Foundation.

\appendix
\section{The autonomous system}\label{appendix}
We use the constraint equation to solve for $A$ in terms of the pressure of the doubly coupled matter field. This gives two branches of solutions for $A$. After substituting the solution for $A$ into the Friedmann equation, we solve it for $\rho_\chi$ and obtain 
\begin{eqnarray}
\rho_\chi=((2m^2\mpl^2\kappa_3-2\alpha\beta^3P_\chi)^3(-3H^2-m^2(1+\frac{m\mpl (\bar{P}_\chi+m\mpl \kappa_2)-2\alpha^2\beta^2P_\chi}{2m^2\mpl^2\kappa_3-2\alpha\beta^3P_\chi})\nonumber\\
(\kappa_1+\frac{3\kappa_2}{2}+3\kappa_3+((m\mpl (\bar{P}_\chi+m\mpl \kappa_2)-2\alpha^2\beta^2P_\chi)(m\mpl \kappa_3(\bar{P}_\chi -2m\mpl \nonumber\\
(\kappa_2+\kappa_3))+\alpha\beta^2(3\beta\kappa_2-2\alpha\kappa_3+2\beta\kappa_3)P_\chi))/(4(m^2\mpl^2\kappa_3-\alpha\beta^3P_\chi)^2))+6H^2\Omega_\phi))\nonumber\\
/(m^3\mpl\alpha(\bar{P}_\chi\beta+m\mpl(\beta\kappa_2-2\alpha\kappa_3))^3) \,.
\end{eqnarray}
Using the accelerating equation and replacing $\rho_\chi$ by the above expression, we write $\dot{H}$ in terms of $H$ and $P_\chi$
\begin{eqnarray}
\dot{H}=\frac{1}{8(m^2\mpl^2\kappa_3-\alpha\beta^3P_\chi)^2}(-12H^2(m^2\mpl^2\kappa_3-\alpha\beta^3P_\chi)^2+m^2(-2\alpha^2\beta^3\nonumber\\
(3\beta(\alpha+\beta)(-\alpha\kappa_2+\beta(2\kappa_1+\kappa_2))+2(\alpha^3+\beta^3)\kappa_3)P_\chi^2+2m\mpl\alpha P_\chi   \nonumber\\
(m\mpl(3\alpha^2\beta\kappa_2\kappa_3-2\alpha^3\kappa_3^2-3\alpha\beta^2(\kappa_2^2-2\kappa_1\kappa_3)+\beta^3(3\kappa_1\kappa_2+6(2\kappa_1+\kappa_2)\kappa_3\nonumber\\
+4\kappa_3^2))+2\alpha\beta(\beta^2\kappa_1-\alpha\beta\kappa_2+\alpha^2\kappa_3)\bar{P}_\chi)+m^3\mpl^3(m\mpl(\kappa_2+2\kappa_3)(\kappa_2^2-2\kappa_2\kappa_3\nonumber\\
-2\kappa_3(3\kappa_1+\kappa_3))+(\kappa_2^2-4\kappa_1\kappa_3)\bar{P}_\chi)))
\end{eqnarray}
Next, we solve the conservation equation of the standard matter field for $\dot{\Omega}_\phi$ after replacing the expressions for $A$, $\rho_\chi$ and $\dot{H}$
\begin{eqnarray}
\dot{\Omega}_\phi=(m^2(2\alpha^2\beta^3(3\beta(\alpha+\beta)(-\alpha\kappa_2+\beta(2\kappa_1+\kappa_2))+2(\alpha^3+\beta^3)\kappa_3)P_\chi^2+2m\mpl\alpha P_\chi \nonumber\\
(m\mpl(3\beta^2\kappa_2(-\beta \kappa_1+\alpha \kappa_2)-3\beta(2\beta(\alpha+2\beta)\kappa_1+(\alpha^2+2\beta^2)\kappa_2)\kappa_3+2(\alpha^3-2\beta^3)\kappa_3^2) \nonumber\\
+2\beta(-\beta^2\kappa_1+\alpha\beta\kappa_2-\alpha^2\kappa_3)\bar{P}_\chi+m^3\mpl^3(-m\mpl(\kappa_2+2\kappa_3)(\kappa_2^2-2\kappa_2\kappa_3\nonumber\\
-2\kappa_3(3\kappa_1+\kappa_3))-(\kappa_2^2-4\kappa_1\kappa_3)\bar{P}_\chi))\Omega_\phi)/(4H(m^2\mpl^2\kappa_3-\alpha\beta^3P_\chi)^2) \,.
\end{eqnarray}
Finally, the last equation of the autonomous system is the conservation equation of the matter field living on the effective metric, which we solve for $\dot{P}_\chi$ in terms of $H$, $P_\chi$ and $\Omega_\phi$ after having used the above expressions 
\begin{eqnarray}
\dot{P}_\chi= -\frac{\mathcal{Y}}{\mathcal{X}}
\end{eqnarray}
where the numerator is given by 
\begin{eqnarray}
\mathcal{Y}&=&H \bar{P}_\chi(\kappa_3 m^2 M_{\rm Pl}^2-\alpha  \beta
   ^3 P_\chi) (m^2 (-(m^5 M_{\rm Pl}^5
   (m M_{\rm Pl} (8 \kappa_1^2 \kappa_3^2
   -\kappa_2^2 \kappa_3 (7 \kappa_1+3 \kappa_3) \nonumber\\
  && -2 \kappa_2 \kappa_3^2
   (3 \kappa_1+\kappa_3) 
   +\kappa_2^4)+\bar{P}_\chi(-\kappa_2 \kappa_3 (5 \kappa_1+3 \kappa_3)
   -2 \kappa_3^2 (3 \kappa_1+\kappa_3)+\kappa_2^3))  \nonumber\\
  && +\alpha  m^3 M_{\rm Pl}^3 P_\chi
   (m M_{\rm Pl} (-2 \alpha ^3 \kappa_2
   \kappa_3^2
   +\alpha ^2 \beta  \kappa_3 (7
   \kappa_2^2-16 \kappa_1 \kappa_3)+2 \alpha  \beta ^2 (\kappa_2
   \kappa_3 (14 \kappa_1+3 \kappa_3)  \nonumber\\
   &&+2 \kappa_3^2 (3 \kappa_1+\kappa_3)-4
   \kappa_2^3)+\beta ^3 (-16 \kappa_1^2 \kappa_3
   +\kappa_1 \kappa_2 (7 \kappa_2+12 \kappa_3)
   +2 \kappa_2 
   \kappa_3 (3 \kappa_2+2 \kappa_3)))  \nonumber\\
  && +\bar{P}_\chi(-2 \alpha ^3 \kappa_3^2
   +5 \alpha ^2 \beta  \kappa_2 \kappa_3 
   +2 \alpha  \beta ^2 (5 \kappa_1 \kappa_3-3
   \kappa_2^2) +\beta ^3 (6 \kappa_3 (2
   \kappa_1+\kappa_2)+5 \kappa_1
   \kappa_2+4 \kappa_3^2)))  \nonumber\\
 &&  -\alpha ^2 \beta ^2 m M_{\rm Pl}
   P_\chi^2 (m M_{\rm Pl} (-12 \alpha ^4
   \kappa_3^2+24 \alpha ^3 \beta  \kappa_2
   \kappa_3
   -\alpha ^2 \beta ^2 (4 \kappa_1
   \kappa_3  
   +17 \kappa_2^2)   +4 \alpha  \beta
   ^3 (7 \kappa_1 \kappa_2 \nonumber\\
   &&+6 \kappa_1 \kappa_3 
   +3 \kappa_2 \kappa_3+2
   \kappa_3^2)+\beta ^4 (-8 \kappa_1^2+6 \kappa_1 \kappa_2   
   +3 \kappa_2^2+2 \kappa_2 \kappa_3))
   +\beta \bar{P}_\chi(6 \alpha ^3 \kappa_3
   -7 \alpha ^2 \beta \kappa_2  \nonumber\\
  && +10 \alpha  \beta ^2 \kappa_1 
   +\beta ^3  (6 \kappa_1+3 \kappa_2+2 \kappa_3)))
   +2 \alpha ^4 \beta ^5 P_\chi^3
   (2 \kappa_3 (\alpha ^3+\beta ^3) 
   +3 \beta  (\alpha +\beta ) (\beta  (2 \kappa_1+\kappa_2)  \nonumber\\
  && -\alpha  \kappa_2))))    
   -6 H^2 (2 \Omega_\phi-1)
   (\alpha  \beta ^3 P_\chi-\kappa_3 m^2
   M_{\rm Pl}^2)^2 (m M_{\rm Pl} (\kappa_2 m
   M_{\rm Pl}+\bar{P}_\chi) 
   -2 \alpha ^2 \beta ^2 P_\chi))
\end{eqnarray}
and the denominator is given by
\begin{align}
\mathcal{X}&=
m M_{\rm Pl} \alpha  \beta  ((-2 P_\chi^3 \alpha
   ^3 (\alpha +\beta ) (\kappa_3 \alpha ^2-\beta 
   \kappa_2 \alpha +\beta ^2 \kappa_1)
   (3 \beta  (\beta  (2 \kappa_1+\kappa_2)-\alpha  \kappa_2)+2 (\alpha ^2 \nonumber\\
  & -\beta  \alpha
   +\beta ^2) \kappa_3) \beta ^6+m M_{\rm Pl}
   P_\chi^2 \alpha ^2 (\bar{P}_\chi\beta  (8
   \kappa_3^2 \alpha ^4-16 \beta  \kappa_2
   \kappa_3 \alpha ^3+\beta ^2 (7 \kappa_2^2+20 \kappa_1 \kappa_3) \alpha ^2 \nonumber\\
 &  +2 
   \beta ^3 (2 \kappa_3^2+6 \kappa_1
   \kappa_3+3 \kappa_2 \kappa_3-7
   \kappa_1 \kappa_2) \alpha +\beta ^4
   (4 \kappa_1^2-6 \kappa_2 \kappa_1-3 \kappa_2^2-2 \kappa_2 \kappa_3)) \nonumber\\
&   +m M_{\rm Pl} (-24 \kappa_3^3
   \alpha ^5+60 \beta  \kappa_2 \kappa_3^2 \alpha
   ^4-4 \beta ^2 \kappa_3 (13 \kappa_2^2+8
   \kappa_1 \kappa_3) \alpha ^3+\beta ^3
   (17 \kappa_2^3+6 \kappa_3 (9
   \kappa_1 \nonumber\\
  & +\kappa_3) \kappa_2+4
   \kappa_3^2 (3 \kappa_1+\kappa_3)) \alpha ^2+2 \beta ^4 (4 \kappa_3
   \kappa_1^2-\kappa_2 (17 \kappa_2+6
   \kappa_3) \kappa_1 \nonumber\\
 &  -\kappa_2
   \kappa_3 (3 \kappa_2+2 \kappa_3)) \alpha +\beta ^5 (2 (7 \kappa_2+18
   \kappa_3) \kappa_1^2-6 (\kappa_2^2-3 \kappa_3 \kappa_2-2 \kappa_3^2) \kappa_1-\kappa_2^2 (3
   \kappa_2 \nonumber\\
 &  +2 \kappa_3))))
   \beta ^3+m^5 M_{\rm Pl}^5 (\bar{P}_\chi
   ((\kappa_2^4-3 \kappa_3 (2
   \kappa_1+\kappa_3) \kappa_2^2-2
   \kappa_3^2 (3 \kappa_1+\kappa_3)
   \kappa_2 \nonumber\\
  & +4 \kappa_1^2 \kappa_3^2) \beta ^2+2 \alpha  \kappa_3
   (-\kappa_2^3+\kappa_3 (5 \kappa_1+3 \kappa_3) \kappa_2+2 \kappa_3^2
   (3 \kappa_1+\kappa_3)) \beta  \nonumber\\
  & +\alpha ^2
   \kappa_3^2 (\kappa_2^2-4 \kappa_1 \kappa_3))+m M_{\rm Pl}
   ((\kappa_2^5-\kappa_3 (8
   \kappa_1+3 \kappa_3) \kappa_2^3-2
   \kappa_3^2 (3 \kappa_1+\kappa_3)
   \kappa_2^2 \nonumber\\
  & +2 \kappa_1 \kappa_3^2 (7
   \kappa_1+3 \kappa_3) \kappa_2+4
   \kappa_1 \kappa_3^3 (3\kappa_1+\kappa_3)) \beta ^2+2 \alpha  \kappa_3 (-\kappa_2^4 \nonumber\\
&   +\kappa_3 (7
   \kappa_1+3 \kappa_3) \kappa_2^2+2
   \kappa_3^2 (3 \kappa_1+\kappa_3)
   \kappa_2-8 \kappa_1^2 \kappa_3^2) \beta +\alpha ^2 \kappa_3^2 (\kappa_2+2 \kappa_3) (\kappa_2^2 \nonumber\\
  & -2
   \kappa_3 \kappa_2-2 \kappa_3 (3
   \kappa_1+\kappa_3))))+2 m^3
   M_{\rm Pl}^3 P_\chi \alpha  (\bar{P}_\chi\beta 
   (4 \kappa_3^3 \alpha ^4-8 \beta  \kappa_2
   \kappa_3^2 \alpha ^3+8 \beta ^2 \kappa_3
   (\kappa_2^2 \nonumber\\
&   -\kappa_1 \kappa_3) \alpha ^2-\beta ^3 (3 \kappa_2^3-2
   \kappa_1 \kappa_3 \kappa_2+4
   \kappa_3^3+6 (2 \kappa_1+\kappa_2)
   \kappa_3^2) \alpha  \nonumber\\
&   +\beta ^4 (-4
   \kappa_3 \kappa_1^2+3 \kappa_2
   (\kappa_2+2 \kappa_3) \kappa_1+\kappa_2 \kappa_3 (3 \kappa_2+2
   \kappa_3)))+m M_{\rm Pl} (-2
  \kappa_3^4 \alpha ^5 \nonumber\\
  &  +5 \beta  \kappa_2
   \kappa_3^3 \alpha ^4+\beta ^2 \kappa_3^2
   (24 \kappa_1 \kappa_3-11 \kappa_2^2) \alpha ^3+\beta ^3 \kappa_3 (11
   \kappa_2^3+3 \kappa_3 (\kappa_3 \nonumber\\
 &  -11
   \kappa_1) \kappa_2+2 \kappa_3^2 (3
   \kappa_1+\kappa_3)) \alpha ^2-\beta ^4
   (4 \kappa_2^4+\kappa_3 (3 \kappa_3-10 \kappa_1) \kappa_2^2+2 \kappa_3^2 (3 \kappa_1+\kappa_3) \kappa_2 \nonumber\\
 &  -10 \kappa_1^2 \kappa_3^2) \alpha
   +\beta ^5 (-2 \kappa_3 (7 \kappa_2+9
   \kappa_3) \kappa_1^2+(4 \kappa_2^3+6 \kappa_3 \kappa_2^2-9 \kappa_3^2 \kappa_2-6 \kappa_3^3)
   \kappa_1 \nonumber\\
 &  +\kappa_2^2 \kappa_3 (3
   \kappa_2+2 \kappa_3))))) m^2+6 H^2 (P_\chi
   \alpha  \beta ^3-m^2 M_{\rm Pl}^2 \kappa_3)^2
   (2 P_\chi \alpha  (\kappa_3 \alpha
   ^2 \nonumber\\
&   -\beta  \kappa_2 \alpha +\beta ^2 \kappa_1) \beta ^3+m M_{\rm Pl} (\kappa_2
   (\text{tp}+m M_{\rm Pl} \kappa_2) \beta ^2-2 (\bar{P}_\chi
   \alpha +m M_{\rm Pl} (\beta  \kappa_1+\alpha 
   \kappa_2)) \kappa_3 \beta \nonumber\\
 &  +2 m M_{\rm Pl}
   \alpha ^2 \kappa_3^2)) (2 \Omega_\phi-1))
\end{align}

	\bibliographystyle{JHEPmodplain}
	\bibliography{cosmology_MG_geff_matter}

\end{document}